\documentclass[12pt,a4paper]{article}
\pdfoutput=1

\usepackage{macros}

\preprint{}
\title{Instanton counting and O-vertex}

\author[a]{Satoshi Nawata}
\author[b]{Rui-Dong Zhu}

\affiliation[a]{Department of Physics and Center for Field Theory and Particle Physics, Fudan University, 220, Handan Road, 200433 Shanghai, China}
\affiliation[b]{Institute for Advanced Study \& School of Physical Science and Technology,\\ Soochow University, Suzhou 215006, China
}
\emailAdd{snawata@gmail.com}
\emailAdd{rdzhu@suda.edu.cn}
\abstract{We present closed-form expressions of unrefined instanton partition functions for gauge groups of type $BCD$ as sums over Young diagrams. For $\SO(n)$ gauge groups, we provide a fivebrane web picture of our formula based on the vertex-operator formalism of the topological vertex with a new type called O-vertex for an O5-plane. }

\begin{document}

\allowdisplaybreaks

\maketitle
\section{Introduction}

At the dawn of the twentieth century, a Young diagram was introduced to represent a partition of a positive integer. It is a basic tool in physics to ``count states''. The harmonic oscillator in quantum mechanics is a salient example in which states can be counted by Young diagrams. In mathematics, it was first used in representation theory, and it gradually became versatile in other fields due to its fundamental nature. From the modern viewpoint, it plays the role of bridges that connect different areas.

The seminal paper \cite{Nekrasov:2002qd} considers 4d and 5d gauge theories with eight supercharges on the $\Omega$-background and employs equivariant localizations on instanton moduli spaces. An exact instanton partition function reproduces non-perturbative dynamics of a 4d $\cN=2$ theory \cite{Seiberg:1994rs,Seiberg:1994aj}. Moreover, for a gauge group of type $A$, fixed points of the equivariant action are classified by ordered sets of Young diagrams and the computation of instanton partition functions significantly simplifies to combinatorics of Young diagrams.  Consequently, exact results \cite{Nekrasov:2002qd,Pestun:2007rz} in 4d and 5d theories with eight supercharges connect instantons to \emph{BPS/CFT correspondence} in physics (for instance \cite{Alday:2009aq,Nekrasov:2012xe,NPS,BPS/CFT,Kimura-Pestun}), and \emph{geometric representation theory} in mathematics (for instance \cite{Braverman:2004vv,Schiffmann:2012aa,Maulik:2012wi,Braverman:2014xca}).

A 5d theory on the $\Omega$-background effectively reduces to supersymmetric quantum mechanics on instanton moduli spaces.
For gauge groups of type $BCD$, an instanton moduli space admits an ADHM description \cite{Atiyah:1978ri}, which is a Nakajima quiver variety \cite{nakajima1994instantons} of Jordan type \cite{Benvenuti:2010pq}. Using the ADHM descriptions, instanton partition functions are successfully expressed as Jeffrey-Kirwan contour integrals in outstanding works \cite{Nekrasov-Shadchin,Kim:2012gu}.
Moreover, $\SO(7)$ and $G_2$ instanton partition functions are written by sums over Young diagrams in \cite{Kim:2018gjo}, employing ingenious representation theoretic methods. Nonetheless, unlike gauge groups of type $A$, it has been an open problem for a long time to classify poles and obtain closed-form expressions as residue sums from the ADHM descriptions for gauge groups of type $BCD$. This paper provides a solution for $\SO(n)$ gauge groups and a conjectural expression for $\Sp(N)$ gauge groups to the long-standing problem at the \emph{unrefined} level in \S\ref{sec:ADHM}.

String theory provides other perspectives for instanton partition functions. For a gauge group of type $A$, the topological vertex \cite{AKMV} computes a partition function from a toric Calabi-Yau three-fold thanks to geometric engineering \cite{Katz:1996fh,Katz:1997eq}. Also, the duality \cite{Leung-Vafa} between a toric Calabi-Yau three-fold and a $(p,q)$-fivebrane web allows us to identify it with the partition function of a 5d gauge theory on an intersection of fivebranes where instantons are realized by D1-branes on D5-branes. A 5d gauge theory with a gauge group of type $BCD$ is realized on a stack of $N$ D5-branes on an O5-plane in string theory. Recently, an analog of the topological vertex for an O5-plane is proposed in \cite{Hayashi:2020hhb}, called an \emph{O-vertex}. Using the O-vertex, we obtain a closed-form expression of a partition function of an $\SO(n)$ gauge theory in \S\ref{sec:O}. Remarkably, it coincides with the expression obtained from the ADHM description. At the end of \S\ref{sec:O}, we also show that the O-vertx can provide a closed-form expression of the $G_2$ instanton partition function by using fivebrane web diagrams \cite{G-type,Hayashi:2018lyv}.

This paper aims to present unrefined instanton partition functions for gauge groups of type $BCD$ as concisely as possible. Since Jeffrey-Kirwan contour integrals from the ADHM descriptions are rather complicated, we delegate the pole analysis to Appendix \ref{app:ADHM}. Also, detailed computations for correlation functions involving the O-vertex are given in Appendix \ref{app:M-proof}.  In Appendix \ref{app:4d}, we give expressions of 4d unrefined instanton partition functions for gauge groups of type $BCD$. We believe that we use the standard notations in literature. Nevertheless, the reader can refer to Appendix \ref{app:notations} for notations and definitions necessary for this paper.

\section{Instantons from ADHM}\label{sec:ADHM}
\subsection{\texorpdfstring{$\SO(n)$}{SO(n)} instanton partition functions}\label{sec:SO-ADHM}

A 5d theory on the $\Omega$-background effectively reduces to supersymmetric quantum mechanics on the instanton moduli spaces.  Fortunately, the $k$-instanton moduli space for $\SO(n)$ gauge group admits a standard ADHM formulation. Hence, the supersymmetric quantum mechanics on the $k$-instanton moduli space is described by the $\Sp(k)$ gauge theory with one hypermultiplet in the second rank antisymmetric representation and $n$ hypermultiplets in the fundamental representation \cite{Nekrasov-Shadchin,Benvenuti:2010pq}.
   The partition function \cite{Nekrasov-Shadchin} of the supersymmetric quantum mechanics is given by the contour integral
\begin{align}\label{SO-contour}
Z^{k}_{\SO(n)}=&\frac{1}{k!2^k}\oint\prod_{I=1}^k\frac{d\phi_I}{2\pi i}\cdot
\frac{\prod\limits_{I\neq J}2\sinh\frac{\phi_{I}-\phi_{J}}{2}\cdot
\prod\limits_{I,J}2\sinh\frac{2\epsilon_+-\phi_{I}+\phi_{J}}{2}}
{\prod\limits_{I,J}2\sinh\frac{\epsilon_{1,2}+\phi_{I}-\phi_{J}}{2}\prod\limits_{I=1}^k2^\chi \sinh^\chi \frac{\epsilon_{+} \pm \phi_{I}}{2}\prod\limits_{s=1}^N 2\sinh\frac{\epsilon_+\pm\phi_I\pm a_s}{2}}
\cr
&\times\frac{\prod\limits_{I\leq J}2\sinh\frac{\pm(\phi_I+\phi_J)}{2}
\cdot 2\sinh\frac{2\epsilon_+\pm(\phi_I+\phi_J)}{2}}
{\prod\limits_{I<J}2\sinh\frac{\epsilon_{1}\pm(\phi_I+\phi_J)}{2}\cdot 2\sinh\frac{\epsilon_{2}\pm(\phi_I+\phi_J)}{2}}\ .
\end{align}
where $n=2 N+\chi$ with $N=\lfloor \frac{n}{2}\rfloor$, $n \equiv \chi\ (\bmod \ 2)$.

Remarkably, in the unrefined limit $\e_1=-\e_2=\hbar$, the JK-poles are classified by $N$-tuples $\vec{\lambda}$ of Young diagrams with the total number of boxes $\sum_{s=1}^N|\lambda^{(s)}|=k$ as analyzed in Appendix \ref{app:ADHM}.
We specify a pole location associated with a content $(i,j)\in \lambda^{(s)}$ as
\begin{equation}\label{phi(s)}
  \phi_{i,j}(s)= a_s+(i-j)\hbar\  .
\end{equation}
By defining a familiar factor \cite{Nekrasov:2002qd}
\begin{equation}\label{Nij}
  N_{i,j}(s,t):= a_s- a_t-\hbar(a_{i,j}(\lambda^{(s)})+l_{i,j}(\lambda^{(t)})+1)\ ,
\end{equation}
the residue sum is then expressed as
\begin{align} \label{SO-ADHM}
  Z^k_{\SO(n)}= &\sum_{\vec{\lambda}}\prod_{s=1}^N\prod_{(i,j)\in \lambda^{(s)}}
  \frac{ 2^{4}\sinh^{4}\phi_{i,j}(s) }
  { {\color{red}{2^{2\chi}\sinh^{2\chi}\frac{\phi_{i,j}(s)}2 }}\prod\limits_{t=1}^N {\color{red}{4\sinh^2\frac{N_{i,j}(s,t)}{2}}}\cdot 4\sinh^2\frac{\phi_{i,j}(s)+ a_t}{2}}\cr
  &\times \prod_{s\leq t}^N\prod_{\substack{(i,j)\in \lambda^{(s)},(m,n)\in \lambda^{(t)}\\ (i,j)< (m,n) }}
  \frac{16\sinh^4\frac{\phi_{i,j}(s)+\phi_{m,n}(t)}{2}}
  { 4\sinh^2\frac{\phi_{i,j}(s)+\phi_{m,n}(t)\pm\hbar}{2}}~.
\end{align}
Here we define a total ordering on boxes of the Young diagrams:
\be \lambda^{(s)}\ni (i,j)< (m,n)\in \lambda^{(t)} \quad \text{if} \quad \begin{cases}s<t  \\ s=t, i<m \\
s=t, i=m, j<n
\end{cases}~.
\ee
The partition function \eqref{SO-ADHM} is invariant under the inverse $A_i \leftrightarrow A_i^{-1}$ of the Coulomb branch parameters.
In the unrefined limit, the set of poles \eqref{phi(s)} are valid even when 5d hypermultiplets in the fundamental representation of $\SO(N)$ \cite[(4.14)]{Shadchin:2005mx} are included.

As we will see in \S\ref{sec:O}, the formula admits an interpretation of the extended topological vertex formalism with an O5-plane. In fact, the terms colored by red in (\ref{SO-ADHM}) correspond to the Nekrasov factors \eqref{Nekfactor} that appear in the $\U(N)$ instanton partition functions. On the other hand, the other black part corresponds to the contribution from O5-plane, and we call it the $M$-factor \eqref{product}. Comparing with \eqref{SOeven-TV} and \eqref{SOodd-TV} from the topological vertex, we observe that we can rewrite \eqref{SO-ADHM} as
\begin{align}\label{SO-ADHM2}
  &Z^k_{\SO(n)}=\cr= &\sum_{\vec{\lambda}}\prod_{s=1}^N\prod_{(i,j)\in\lambda^{(s)}} \frac{4\sinh^2\frac{2 a_s+\hbar (i-j+(\lambda^{(s)})_{j}^\vee-\lambda_{i}^{(s)})}{2}}{2^{2\chi}\sinh^{2\chi}\frac{\phi_{i,j}(s)}2 \prod\limits_{t=1}^N 4\sinh^2\frac{N_{i,j}(s,t)}{2}} \\
  &\prod_{1\le s< t\le N}\frac{1}{\prod\limits_{(i,j)\in\lambda^{(s)}}4\sinh^2\frac{ a_s+ a_t+\hbar(i+j-1-\lambda^{(s)}_{i}-\lambda^{(t)}_{j})}{2} \prod\limits_{(m,n)\in\lambda^{(t)}}4\sinh^2\frac{ a_s+ a_t+\hbar (1-m-n+(\lambda^{(t)})^\vee_{n}+(\lambda^{(s)})^\vee_{m})}{2}}~.\nonumber
\end{align}

The formulas  \eqref{SO-ADHM} and \eqref{SO-ADHM2} allow us to check the Lie algebra theoretic relations among instanton partition functions.
For instance, the isomorphism $\mathfrak{so}(4)\simeq \mathfrak{su}(2)\oplus\mathfrak{su}(2)$ indicates the identity of the partition functions
$$
  Z_{\SO(4)}(A_1,A_2)=Z_{\SU(2)}(A=A_1^{\frac12} A_2^{\frac12})Z_{\SU(2)}(A=A_1^{\frac12} A_2^{-\frac12})Z_{\U(1)}~,
$$
where
$$
Z_{\U(1)}=\exp\Bigl(\sum_{k=1}^\infty\frac{\frakq^kq^k}{k(1-q^k)^2}\Bigr)~.
$$
Here $\frakq$ is the instanton counting parameter, and we confirm this identity up to 10-instanton.
The isomorphism $\mathfrak{so}(6)\simeq\mathfrak{su}(4)$ of Lie algebras leads to the equality of the partition functions $$Z_{\SO(6)}(A_1,A_2,A_3)=Z_{\SU(4)}(A_1,A_2,A_3)~,$$ which we verify up to 6-instanton. Furthermore, using an embedding $\SU(4)\subset \SO(7)$, the $\SO(7)$ instanton partition function has been written as a sum over 4-tuples of Young diagrams in \cite[(2.18)]{Kim:2018gjo}. Although the expression \eqref{SO-ADHM} for $\SO(7)$ is written in terms of a sum over 3-tuples of Young diagrams, we check that it agrees with  the unrefined limit of \cite[(2.18)]{Kim:2018gjo} up to 6-instanton.

\subsection{\texorpdfstring{$\Sp(N)$}{Sp(N)} instanton partition functions}\label{sec:Sp-ADHM}

Similarly, the $\Sp(N)$ instanton moduli space can be also described by the ADHM quiver.
Supersymmetric quantum mechanics on the $k$-instanton moduli space is described by the $\OO(k)$ gauge theory with one hypermultiplet in the second rank symmetric representation and $N$ hypermultiplets in the fundamental representation \cite{Nekrasov-Shadchin,Benvenuti:2010pq}. Since $\OO(k)$ has two connected components $\OO(k)_\pm$ ($\det =\pm$), the instanton partition function receives contribution $Z^{k}_\pm$ from each connected component $\OO(k)_\pm$. A choice of taking the sum or difference of $Z^{k}_\pm$ corresponds to the one of the discrete $\theta$-angle that originates from $\pi_4(\Sp(N))=\mathbb{Z}_2$ \cite{Bergman:2013ala}. Thus,  the $k$-instanton partition function with trivial $\theta$-angle is given by
\begin{equation}
  Z^k_{\theta=0}=\frac{Z^k_++ Z^k_-}{2}\ ,
\end{equation}
whereas that with non-trivial $\mathbb{Z}_2$ element is given by
\begin{equation}
  Z^k_{\theta=\pi}=(-1)^k\frac{Z^k_+-Z^k_-}{2}\ .
\end{equation}

Using the description of supersymmetric quantum mechanics on the ADHM description, the contour integral expressions of $\Sp(N)$ instanton partition functions are given in \cite{Kim:2012gu,Kim:2012qf,Hwang:2014uwa}. The formulas are rather involved so that we summarize the integral expressions in Appendix \ref{app:ADHM}. In the unrefined limit, the JK-residues of $Z^{k}_\pm$ are classified by $(N+4)$-tuples of Young diagrams with the total number of boxes $\sum_{s=1}^{N+4}|\lambda^{(s)}|=\ell$. The relation between the instanton number $k$ and the total number $\ell$ of boxes will be given below. More remarkably, they can be expressed by a single universal formula
\begin{align} \label{Sp-ADHM}
  \wt Z_{\Sp(N)}^\ell(A_1,\ldots,A_{N+4};q)= &\sum_{\vec{\lambda}}C_{\vec{\lambda},\vec{A}}\prod_{s=1}^{N+4}\prod_{(i,j)\in \lambda^{(s)}}
  \frac{2^{4}\sinh^{4}\phi_{i,j}(s)}
  {\prod\limits_{t=1}^{N+4} 4\sinh^2\frac{N_{i,j}(s,t)}{2}\cdot 4\sinh^2\frac{\phi_{i,j}(s)+ a_t}{2}}\cr
  &\times \prod_{s\leq t}^{N+4}\prod_{\substack{(i,j)\in \lambda^{(s)},(m,n)\in \lambda^{(t)}\\ (i,j)< (m,n) }}
  \frac{16\sinh^4\frac{\phi_{i,j}(s)+\phi_{m,n}(t)}{2}}
  { 4\sinh^2\frac{\phi_{i,j}(s)+\phi_{m,n}(t)\pm\hbar}{2}},
\end{align}
where $\phi_{i,j}(s)$ and $N_{i,j}(s,t)$ are given in \eqref{phi(s)} and \eqref{Nij}, and $C_{\vec{\lambda},\vec{A}}$ is a constant
\be\label{weight}
C_{\vec{\lambda},\vec{A}}=\prod_{s=N+1}^{N+4}C_{\lambda^{(s)},A_s}~.
\ee
Each constant is defined as follows: $C_{\lambda^{(s)}=\emptyset,A_s}=1$, and
\begin{align}\label{weight2}
C_{\lambda^{(s)},A_s=\pm1,\pm q^{\frac12}}&=\frac{2^{2m-1}}{\binom{2m-1}{m-1}}\qquad  \textrm{  where $m$ is the number of rows with $\lambda_j^{(s)}\ge j$}~,\cr
C_{\lambda^{(s)},A_s=\pm q}&=\frac{2^{2m}}{\binom{2m+1}{m}}\qquad  \textrm{  where $m$ is the number of rows with $\lambda_j^{(s)}\ge j+1$}~.
\end{align}
(See also \eqref{const1} and \eqref{const2} for more illustrative description.)
These constants are in principle determined by the number of poles and the determinant of JK vectors that define a cone, and we \emph{conjecture} \eqref{weight2} by comparing \eqref{Sp-ADHM} and JK-residues of \eqref{Sp-contour}.
The formula \eqref{Sp-ADHM} is a close cousin of that of $\SO(2N)$ instanton partition function \eqref{SO-ADHM}. The difference is that the non-trivial constant \eqref{weight} is involved, and poles of four other types ($s=N+1,\ldots,N+4$) do not involve the Coulomb branch parameters. Note that \eqref{Sp-ADHM} can be written as in the form of \eqref{SO-ADHM2}.

Performing the JK residues as in Appendix \ref{app:ADHM}, $Z^k_\pm$ can be written as
\begin{align}\label{Sp-ADHM-pm}
Z^{k=2\ell}_+=& 2 \wt Z_{\Sp(N)}^{\ell}(A_1,\ldots,A_N,A_{N+1}=q^{\frac12},A_{N+2}=-q^{\frac12},A_{N+3}=1,A_{N+4}=-1;q) \cr
Z^{k=2\ell+1}_+=&\frac{1}{4 \sinh^2 \frac{\hbar}{2} \prod\limits_{i=1}^{N} 4 \sinh^2 \frac{  a_{i}}{2}}\cr
  &  \wt Z_{\Sp(N)}^{\ell}(A_1,\ldots,A_N,A_{N+1}=q^{\frac12},A_{N+2}=-q^{\frac12},A_{N+3}=q,A_{N+4}=-1;q)  \cr
Z^{k=2\ell}_-= &2(-1)^{N}\frac{1}{4 \sinh^2 \frac{\hbar}{2}\cdot 4 \sinh^2 \hbar \prod\limits_{i=1}^{N} 4 \sinh^2\left( a_{i}\right)} \cr
    & \wt Z_{\Sp(N)}^{\ell-1}(A_1,\ldots,A_N,A_{N+1}=q^{\frac12},A_{N+2}=-q^{\frac12},A_{N+3}=q,A_{N+4}=-q;q) \cr
Z^{k=2\ell+1}_-=& (-1)^{N}\frac{1}{4 \sinh^2 \frac{\hbar}{2} \prod\limits_{i=1}^{N} 4 \cosh^2 \frac{  a_{i}}{2}} \cr
  & \wt Z_{\Sp(N)}^{\ell}(A_1,\ldots,A_N,A_{N+1}=q^{\frac12},A_{N+2}=-q^{\frac12},A_{N+3}=1,A_{N+4}=-q;q)~.
\end{align}
The pole structure will remain as it is even if we include 5d hypermultiplets in the fundamental representations \cite[(3.48)--(3.50)]{Hwang:2014uwa}.

Again, we can check the Lie algebra theoretic relations of instanton partition functions.
The isomorphisms $\mathfrak{sp}(1)\simeq\mathfrak{su}(2)$ and $\mathfrak{sp}(2)\simeq\mathfrak{so}(5)$  of Lie algebras lead to the equality of the partition functions
\bea\nonumber
Z_{\Sp(1)}(A_1)_{\theta=0}=&Z_{\SU(2)}(A_1)~,\cr
Z_{\Sp(2)}(A_1=Q_1Q_2,A_2=Q_2)_{\theta=0}=&Z_{\SO(5)}(A_1=Q_1,A_2=Q_1Q_2^2) ~.
\eea
We check the first and second equality up to 27-instanton and 16-instanton, respectively. The first equality is further checked up to  28-instanton in the 4d limit. 

We can also use string duality to check the formulas.
The 5d $\Sp(N)$ pure gauge theory can be realized by a stack of $N$ D5-branes with an O7${}^-$-plane. By splitting an O7${}^-$-plane into two $(p,q)$-sevenbranes \cite{Sen:1996vd} and performing Hanany-Witten transitions \cite{Hanany:1996ie}, one can obtain a fivebrane diagram \cite[Figure 14]{Hayashi:2016jak} from which the partition function is computed by the topological vertex. The resulting formula is expressed as a sum over $2N$-tuples of Young diagrams with $k$ the total number of boxes. We verify the agreement up to 5-instanton for $\Sp(N)$ $(N<7)$ with both trivial and non-trivial $\theta$-angle. Thus, the string duality yields the highly non-trivial identities for the $\Sp(N)$ instanton partition function.

\section{Instantons from O-vertex}\label{sec:O}

Topological string theory provides an alternative and powerful way to compute instanton partition functions of 5d $\cN=1$ gauge theories. In the case of quiver gauge theories with gauge groups of type $A$, a chain of string dualities takes the gauge theory to the A-model of topological string theory defined on toric Calabi-Yau manifold \cite{Katz:1996fh,Katz:1997eq,Leung-Vafa}. The topological vertex formalism \cite{AKMV,Awata:2005fa,IKV} expresses partition functions in terms of sums over Young diagrams, which strikingly simplifies the evaluations of the partition functions. Recently, inspired by the work of \cite{Kim-Yagi} and its S-dual configuration \cite{D-type}, an extended version of topological vertex formalism was developed for 5d gauge theories with $\SO(n)$ gauge groups \cite{Hayashi:2020hhb} in which a new topological vertex, called the O-vertex, was introduced for O5${}^-$-plane. In this section, we obtain a closed-form analytic expression of the instanton partition function of the 5d $\SO(n)$ pure gauge theory in terms of Young diagrams from the extended topological vertex formalism. The results are consistent with the expressions obtained from the ADHM descriptions in the previous section.
As a by-product, we also present a closed-form expression of the $G_2$ instanton partition function by using the O-vertex.

\subsection{\texorpdfstring{$\SO(2N)$}{SO(2N)} instantons}

The brane construction of the 5d $\cN=1$ SO($2N$) pure gauge theory is well-known in string theory: one can put $N$ D5-branes on the top of O5${}^-$ orientifold (see (\ref{o-vert-soeven})).

\begin{align}
\begin{tikzpicture}
\draw [dashed] (-2,0)--(6,0);
\draw (0,0)--(1,0.5);
\draw (1,0.5)--(3,0.5);
\draw (3,0.5)--(4,0);
\draw (1,0.5)--(1.5,1);
\draw (1.5,1)--(2.5,1);
\draw (2.5,1)--(3,0.5);
\draw (1.5,1)--(1.5,1.5);
\draw (1.5,1.5)--(2.5,1.5);
\draw (2.5,1)--(2.5,1.5);
\draw (1.5,1.5)--(1.3,1.7);
\draw (2.5,1.5)--(2.7,1.7);
\draw[dotted] (1,2) -- (0.2,2.8);
\draw[dotted] (3,2) -- (3.8,2.8);
\draw (0,3)--(4,3);
\draw (-2,3.5)--(6,3.5);
\draw (-2,3.5)--(0,3);
\draw (4,3)--(6,3.5);
\draw (6,3.5)--(7,3.6);
\draw (-2,3.5)--(-3,3.6);
\draw[<->,dotted] (-2,0.05)--(-2,0.45);
\node at (-2,0.25) [left] {$Q_1$};
\draw[<->,dotted] (-2,0.55)--(-2,0.95);
\node at (-2,0.75) [left] {$Q_2$};
\draw[<->,dotted] (-2,1.05)--(-2,1.45);
\node at (-2,1.25) [left] {$Q_3$};
\draw[<->,dotted] (1.55,1.1)--(2.45,1.1);
\node at (2,1) [above] {$Q$};
\draw[<->,dotted] (-2,3.05)--(-2,3.45);
\node at (-2,3.25) [left] {$Q_{N}$};
\node at (2,0) [below] {O5${}^-$};
\node at (5,0) [below] {O5${}^+$};
\node at (-1,0) [below] {O5${}^+$};
\end{tikzpicture}
\label{o-vert-soeven}
\end{align}

It is proposed in \cite{Hayashi:2020hhb} that an extended version of the topological vertex can be applied to the above brane web with orientifold. In this proposal, topological vertices $C_{\mu\nu\lambda}$ are assigned to trivalent vertices as usual, while a new type of topological vertex called O-vertex is attached to the intersection point of the orientifold and (2,1)-brane:
\begin{align}
\begin{tikzpicture}
\draw [dashed] (-2,0)--(2,0);
\draw (0,0)--(1,0.5);
\node at (1,0.5) [above] {$\nu$};
\node at (1,0) [below] {O5${}^-$};
\node at (-1,0) [below] {O5${}^+$};
\node at (0,0.5) {$V_\nu$};
\end{tikzpicture}
\label{o-vert}
\end{align}
where $V_\nu$ is defined as the following entry
\ba
V_\nu(-P)^{|\nu|}=\bra{0}\mathbb{O}(P,q)\ket{\nu},
\ea
of the vertex operator,
\begin{equation}\label{Overtex}
\mathbb{O}(P, q)=\exp (\sum_{n=1}^{\infty}-\frac{P^{2 n}(1+q^{n})}{2 n(1-q^{n})} J_{2 n}+\frac{P^{2 n}}{2 n} J_{n} J_{n}).
\end{equation}
Here, we use the vertex-operator formalism of the topological vertex (see Appendix \ref{app:notations}), and $J_n$ are the modes of the free boson.
The intersection point in the following diagram,
\begin{align}
\begin{tikzpicture}[xscale=-1]
\draw [dashed] (-2,0)--(2,0);
\draw (0,0)--(1,0.5);
\node at (1,0.5) [above] {$\nu$};
\node at (0,0.5) {$W_\nu$};
\node at (1,0) [below] {O5${}^-$};
\node at (-1,0) [below] {O5${}^+$};
\end{tikzpicture}
\label{o-vert-w}
\end{align}
is assigned the dual O-vertex,
\ba
W_\nu(-P)^{|\nu|}q^{\frac{\kappa(\nu^\vee)}{2}}=\bra{0}\mathbb{O}(P,q)\ket{\nu^\vee}.
\ea

By applying the extended topological vertex formalism to the web diagram (\ref{o-vert-soeven}), the partition function  can be written as
\begin{align}
Z_{\SO(2N)}^{\textrm{top}}
=&\sum_{\vec{\lambda},\vec{\nu},\vec{\mu}} (-Q)^{\sum_{s=1}^N |\lambda^{(s)}|} (-Q_1)^{|\nu^{(1)}|+|\mu^{(1)}|}(-Q_2)^{|\nu^{(2)}|+|\mu^{(2)}|+|2\lambda^{(1)}|} V_{\nu^{(1)}} W_{\mu^{(1)}}   \prod_{t=2}^N f_{\nu^{(t)}}^{-1}f_{\mu^{(t)}} \cr
& \times \prod_{r=3}^{N} (-Q_r)^{|\nu^{(r)}|+|\mu^{(r)}|}\prod_{s=3}^N\prod_{r=3}^{s} Q_r^{|2(r-3)\lambda^{(s)}|}   \prod_{s=1}^N  f_{\lambda^{(s)}}^{5-2s} C_{(\nu^{(s)})^\vee \nu^{(s+1)} (\lambda^{(s)})^\vee } C_{(\mu^{(s)})^\vee \lambda^{(s)} \mu^{(s+1)}},\nonumber
\end{align}
where the summation is performed over $(3N)$-tuples Young diagrams $(\vec{\lambda},\vec{\nu},\vec{\mu})=(\lambda^{(s)},\nu^{(s)},\mu^{(s)})$ ($s=1,\ldots N$) and we set $\nu^{(N+1)}=\emptyset=\mu^{(N+1)}$.
We expect the above partition function to reproduce the 5d partition function with $\SO(2N)$ gauge group; more precisely, we have
\ba
Z_{\SO(2N)}^{\textrm{top}}=Z_{\SO(2N)}^{\textrm{root}}Z_{\SO(2N)}^{\textrm{inst}},
\ea
where
\ba
Z_G^{\textrm{root}}=\textrm{PE}\lt(\frac{2q}{(1-q)^2}\sum_{\alpha\in\Delta_+}e^{-\alpha\cdot a }\rt),
\ea
is the one-loop factor of the 5d gauge theory, and $Z^{\textrm{inst}}_G$ is the instanton partition function investigated in \S\ref{sec:ADHM}. Note that $Z_{\SO(2N)}^{\textrm{root}}$ can be obtained in the extended topological formalism by setting all $\lambda^{(s)}=\emptyset$.
Summations over $\nu^{(s)},\mu^{(s)}$ can be expressed as a correlation function of vertex operators including the O-vertex \eqref{Overtex}
\begin{align}
&\sum_{\nu^{(s)}}V_{\nu^{(1)}} \prod_{s=1}^N Q_s^{|\nu^{(s)}|}s_{\nu^{(s)}/\nu^{(s+1)}}(q^{-\rho-\lambda^{(s)}})=\bra{0}\mathbb{O}(A_1,q)\prod_{s=1}^N V_-(A_sA_1^{-1} q^{-\rho-\lambda^{(s)}})\ket{0}~.
\end{align}
where $V_-$ is the vertex operator \eqref{VO} for the skew Schur function, and the Coulomb branch parameters are written as $$A_{s}:=\prod_{j=1}^{s} Q_{j}~.$$
This simplifies an expression of the 5d instanton partition function
\begin{align}\label{SOeven-TV}
Z_{\SO(2N)}^{\textrm{inst}}=\frac{Z_{\SO(2N)}^{\textrm{top}}}{Z_{\SO(2N)}^{\textrm{root}}}=&\sum_{\vec{\lambda}}Q^{\sum_{s=1}^N |\lambda^{(s)}|} Q_2^{|2\lambda^{(1)}|}\prod_{s=3}^N\prod_{r=3}^{s} Q_r^{|2(r-3)\lambda^{(s)}|} \prod_{s=1}^N q^{\frac{(5-2s)\kappa(\lambda^{(s)})}{2}} \\
& \qquad  \times s_{\lambda^{(s)}}(q^{-\rho})s_{(\lambda^{(s)})^\vee}(q^{-\rho}) M_{\vec{\lambda}}(\vec{A})^2 \prod_{1\le s< t\le N}  N_{\lambda^{(s)}\lambda^{(t)}}(A_{t}A_s^{-1},q)^{-2}~,\nonumber
\end{align}
where $N_{\mu\nu}$ is the Nekrasov factor \eqref{Nekfactor} and the $M$-factor is defined by
\be\label{M-factor}
M_{\vec{\lambda}}(\vec{A}):=\frac{\bra{0}\mathbb{O}(A_1,q)\prod_{s=1}^N V_-(A_sA_1^{-1} q^{-\rho-\lambda^{(s)}})\ket{0}}{\bra{0}\mathbb{O}(A_1,q)\prod_{s=1}^N V_-(A_sA_1^{-1} q^{-\rho})\ket{0}}~.
\ee
Thus, the $k$-instanton partition function with $\SO(2N)$ gauge group receives contributions from $\vec{\lambda}=(\lambda^{(1)},\ldots,\lambda^{(N)})$ with $k=\sum_{s=1}^N|\lambda^{(s)}|$.

Applying the Baker-Campbell-Hausdorff formula
\be
e^Xe^Y = e^{Y}e^{X-\lt[Y,X\rt]+\frac{1}{2}\lt[Y,\lt[Y,X\rt]\rt]-\frac{1}{3!}\lt[Y,\lt[Y,\lt[Y,X\rt]\rt]\rt]+\dots},
\ee
the $M$-factor can be written as
\be\label{PE}
M_{\vec{\lambda}}(\vec{A})=\text{PE}\Bigl[\frac{qX}{2(1-q)^2}\Bigr]
\ee
where
\begin{align}
X:=&\Bigl[\sum_{s=1}^N  A_s\bigl(q^{\ell(\lambda^{(s)})}+(1-q)\sum_{i=1}^{\ell(\lambda^{(s)})}q^{i-1-\lambda^{(s)}_i}\bigr)\Bigr]^2- \Bigl[\sum_{s=1}^N A_s\Bigr]^2 \cr
&-\Bigl[\sum_{s=1}^N  A_s^2\bigl(q^{2\ell(\lambda^{(s)})}-1+(1-q^2)\sum_{i=1}^{\ell(\lambda^{(s)})}q^{2(i-1-\lambda^{(s)}_i)}\bigr)\Bigr]
\end{align}
One can show that this can be written as
\begin{align}\label{product}
&M_{\vec{\lambda}}(\vec{A})=\\=&\frac{\displaystyle\prod_{s=1}^N\prod_{(i,j)\in\lambda^{(s)}}(1-A_{s}^{2}q^{i-j+(\lambda^{(s)})_{j}^\vee-\lambda_{i}^{(s)})})}
{\displaystyle\prod_{1\le s< t\le N}\prod_{(i,j)\in\lambda^{(s)}}(1-A_{s} A_{t} q^{i+j-1-\lambda^{(s)}_{i}-\lambda^{(t)}_{j}})\prod_{(m,n)\in\lambda^{(t)}}(1-A_{s} A_{t} q^{1-m-n+(\lambda^{(t)})^\vee_{n}+(\lambda^{(s)})^\vee_{m}})}~.\nonumber
\end{align}
The equality between \eqref{PE} and  \eqref{product}  is given in Appendix \ref{app:M-proof}. Note that the $M$-factor is invariant under the permutation group
$$
M_{\lambda^{(1)},\ldots,\lambda^{(N)}}(A_1,\ldots,A_{N})=M_{\lambda^{(\sigma(1))},\ldots,\lambda^{(\sigma(N))}}(A_{\sigma(1)},\ldots,A_{\sigma(N)})
$$
for $\forall \sigma \in \frakS_N$. Plugging the expression \eqref{product} into \eqref{SOeven-TV}, it is equal to \eqref{SO-ADHM2}.

Let us remark that another approach to the $\SO(2N)$ instanton partition functions was proposed in \cite{Ohmori-Hayashi} by the topological vertex with ``trivalent gluing''. Although the method in \cite{Ohmori-Hayashi} has its own advantage (such as applications to gauge groups of other types), it is computationally expensive for $\SO(2N)$ gauge theories. The vertex-operator formalism of the O-vertex in this section provides a closed-form expression \eqref{SOeven-TV} as a sum over $N$-tuples of Young diagrams. Our formula is  computationally efficient so that it would be desirable to find a synergy between our approach and the trivalent gluing in \cite{Ohmori-Hayashi}.

\subsection{\texorpdfstring{$\SO(2N+1)$}{SO(2N+1)} instantons}

In string theory, the 5d $\cN=1$ SO($2N+1$) pure gauge theory is realized by $N$ D5-branes with an $\wt{\textrm{O5}}^-$-plane, which is equivalent to an ${\textrm{O5}}^-$-plane with $\frac12$ D5-brane. The ``effective" brane web of 5d $\SO(2N+1)$ theories that can be used in the extended topological vertex is given in \cite{Zafrir:2015ftn,Bertoldi:2002nn,Hayashi:2020hhb}
\begin{align}
\begin{tikzpicture}
\draw [dashed] (-2,0)--(6,0);
\draw (0,0)--(1,0.5);
\draw (1,0.5)--(2.75,0.5);
\draw (1,0.5)--(1.5,1);
\draw (1.5,1)--(2.75,1);
\draw (1.5,1)--(1.5,1.5);
\draw (1.5,1.5)--(3.25,1.5);
\draw (1.5,1.5)--(1.5-.2,1.5+.2);
\draw (2.75,1)--(3.25,1.5);
\draw (3.25,1.5)--(3.25+.3,1.5+.15);
\draw (3,0.25)--(3.5,0);
\draw (3,0.25)--(2.5,0.25);
\draw (3,0.25)--(2.75,0.5);
\draw (2.75,0.5)--(2.75,1);
\draw[dotted] (1.1,2) -- (0.6,2.8);
\draw[dotted] (3.8,2) -- (4.8,2.8);
\draw (0.5,3)--(5,3);
\draw (-1,3.5)--(7,3.5);
\draw (-1,3.5)--(0.5,3);
\draw (5,3)--(7,3.5);
\draw (7,3.5)--(8.5,3.6);
\draw (-1,3.5)--(-2,3.6);
\node at (2.5,0.25) [left] {$\emptyset$};
\draw[<->,dotted] (3.6,0)--(3.6,0.3);
\node at (3.6,0.3) [right] {$Q_0$};
\draw[<->,dotted] (1.6,1.1)--(2.7,1.1);
\node at (2.2,1.1) [above] {$Q$};
\draw[<->,dotted] (-1.5,0.05)--(-1.5,0.45);
\node at (-1.5,0.25) [left] {$Q_1$};
\draw[<->,dotted] (-1.5,0.55)--(-1.5,0.95);
\node at (-1.5,0.75) [left] {$Q_2$};
\draw[<->,dotted] (-1.5,1.05)--(-1.5,1.45);
\node at (-1.5,1.25) [left] {$Q_3$};
\draw[<->,dotted] (-1.5,3.05)--(-1.5,3.45);
\node at (-1.5,3.25) [left] {$Q_{N}$};
\node at (2,0) [below] {O5${}^-$};
\node at (5,0) [below] {O5${}^+$};
\node at (-1,0) [below] {O5${}^+$};
\end{tikzpicture}
\label{o-vert-soodd}
\end{align}
where we need to take $Q_0\rightarrow 1$ limit at the end of the calculation.
Applying the extended topological vertex to the brane web \eqref{o-vert-soodd}, the partition function can be written as
\begin{align}
&Z_{\SO(2N+1)}^{\textrm{top}}\cr
=&\lim_{Q_0\to 1}\sum_{\substack{\vec{\lambda},\vec{\nu},\vec{\mu}\\|\mu^{(0)}| \leq 2|\mu^{(1)}|}} (-Q)^{\sum_{s=1}^N |\lambda^{(s)}|} (-Q_1)^{|\nu^{(1)}|}(-Q_0)^{|\mu^{(0)}|}(-Q_1/Q_0)^{|\mu^{(1)}|}(-Q_2)^{|\nu^{(2)}|+|\mu^{(2)}|+|\lambda^{(1)}|} \cr
&\qquad \times  V_{\nu^{(1)}} W_{\mu^{(0)}}  C_{(\mu^{(0)})^\vee \mu^{(1)} \emptyset}f_{\mu^{(1)}} \prod_{t=2}^N f_{\nu^{(t)}}^{-1}f_{\mu^{(t)}} \prod_{r=3}^{N} (-Q_r)^{|\nu^{(r)}|+|\mu^{(r)}|}\prod_{s=3}^N\prod_{r=3}^{s} Q_r^{|(2r-5)\lambda^{(s)}|}\cr
&\qquad\times  \prod_{s=1}^N f_{\lambda^{(s)}}^{4-2s} C_{(\nu^{(s)})^\vee \nu^{(s+1)} (\lambda^{(s)})^\vee } C_{(\mu^{(s)})^\vee \lambda^{(s)} \mu^{(s+1)}}\nonumber
\end{align}
where the summation is performed over $(3N+1)$-tuples Young diagrams $(\vec{\lambda},\vec{\nu},\vec{\mu})=(\lambda^{(s)},\nu^{(s)},\mu^{(t)})$ ($s=1,\ldots N$, $t=0,\ldots N$) and we set $\nu^{(N+1)}=\emptyset=\mu^{(N+1)}$.
The summations over $\nu^{(s)},\mu^{(t)}$ can be packaged into the $M$-factors \eqref{M-factor}, and the instanton partition function of 5d $\SO(2N+1)$ theory can be extracted out in the same way as
\begin{align}
&Z_{\SO(2N+1)}^{\textrm{inst}}=\frac{Z_{\SO(2N+1)}^{\textrm{top}}}{Z_{\SO(2N+1)}^{\textrm{root}}}\cr
=&\lim_{Q_0 \to 1}\sum_{\vec{\lambda}}(-Q)^{\sum_{s=1}^N |\lambda^{(s)}|}  \prod_{s=1}^N q^{(2-s)\kappa(\lambda^{(s)})}s_{\lambda^{(s)}}(q^{-\rho})s_{(\lambda^{(s)})^\vee}(q^{-\rho})N_{\emptyset \lambda^{(s)}}(A_s / Q_0, q)^{-1} \cr
&\times Q_2^{|\lambda^{(1)}|}\prod_{s=3}^N\prod_{r=3}^{s} Q_r^{|(2r-5)\lambda^{(s)}|}\prod_{1\le s< t\le N}  N_{\lambda^{(s)}\lambda^{(t)}}(A_tA_{s}^{-1},q)^{-2}~,\cr
&\times  M_{\vec{\lambda}}(\vec{A})M_{\emptyset,\lambda^{(1)},\ldots,\lambda^{(N)}}(Q_0,A_1,\ldots,A_{N})~.\nonumber
\end{align}
It follows from \eqref{product} that the second $M$-factor can be written as
\be\label{M-identity}
M_{\emptyset,\lambda^{(1)},\ldots,\lambda^{(N)}}(Q_0,A_1,\ldots,A_{N})= M_{\lambda^{(1)},\ldots,\lambda^{(N)}}(A_1,\ldots,A_{N})\prod_{s=1}^N N_{\emptyset \lambda^{(s)}}(Q_0  A_s , q)^{-1} ~.
\ee
Thus, we have
\bea\label{SOodd-TV}
Z_{\SO(2N+1)}^{\textrm{inst}}=&\sum_{\vec{\lambda}}(-Q)^{\sum_{s=1}^N |\lambda^{(s)}|}  \prod_{s=1}^N q^{(2-s)\kappa(\lambda^{(s)})}s_{\lambda^{(s)}}(q^{-\rho})s_{(\lambda^{(s)})^\vee}(q^{-\rho})N_{\emptyset \lambda^{(s)}}(A_s , q)^{-2} \cr
&\times M_{\vec{\lambda}}(\vec{A})^2 Q_2^{|\lambda^{(1)}|}\prod_{s=3}^N\prod_{r=3}^{s} Q_r^{|(2r-5)\lambda^{(s)}|}\prod_{1\le s< t\le N}  N_{\lambda^{(s)}\lambda^{(t)}}(A_tA_{s}^{-1},q)^{-2}~~.
\eea
Using the expression \eqref{product}, it is equal to \eqref{SO-ADHM2}.

\subsection{\texorpdfstring{$G_2$}{G2} instantons}
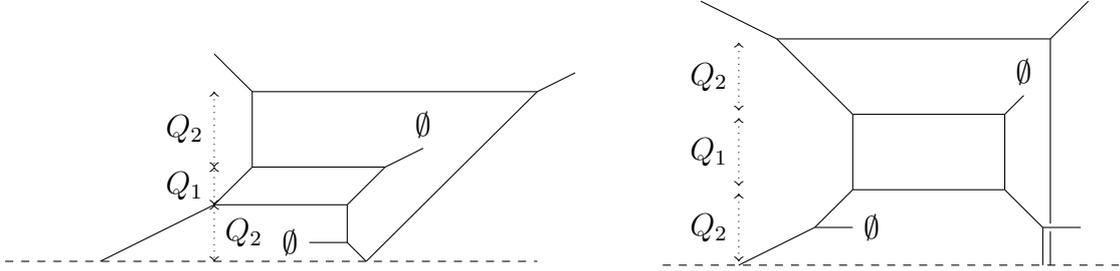
\begin{figure}[ht]
\begin{center}
\begin{tikzpicture}
\draw [dashed] (-2,-0.5)--(5,-0.5);
\draw (-0.75,-0.5)--(0.75,0.25);
\draw (0.75,0.25)--(2.5,0.25);
\draw[<->,dotted] (0.75,0.25)--(0.75,-0.5);
\node at (0.75,-0.125) [right] {$Q_2$};
\draw[<->,dotted] (0.75,0.25)--(0.75,0.75);
\node at (0.75,0.5) [left] {$Q_1$};
\draw[<->,dotted] (0.75,1.75)--(0.75,0.75);
\node at (0.75,1.25) [left] {$Q_2$};
\draw (0.75,0.25)--(1.25,0.75);
\draw (1.25,0.75)--(3,0.75);
\draw (1.25,0.75)--(1.25,1.75);
\draw (1.25,1.75)--(5,1.75);
\draw (1.25,1.75)--(0.75,2.25);
\draw (2.5,0.25)--(2.5,-0.25);
\draw (2,-0.25)--(2.5,-0.25);
\draw (2.5,0.25)--(3,0.75);
\draw (3,0.75)--(3.5,1);
\node at (2,-0.25) [left] {$\emptyset$};
\draw (2.5,-0.25)--(2.75,-0.5);
\draw (2.75,-0.5)--(5,1.75);
\draw (5.5,2)--(5,1.75);
\node at (3.5,1) [above] {$\emptyset$};
\end{tikzpicture}
\hskip 1cm
\begin{tikzpicture}
\draw[dashed] (-1,0)--(5,0);
\draw (0,0)--(1,0.5);
\draw (1,0.5)--(1.5,0.5);
\draw (1,0.5)--(1.5,1);
\draw (1.5,1)--(1.5,2);
\draw (1.5,2)--(0.5,3);
\draw (0.5,3)--(-0.5,3.5);
\draw (1.5,1)--(3.5,1);
\draw (1.5,2)--(3.5,2);
\draw (3.5,1)--(3.5,2);
\draw (3.5,2)--(3.75,2.25);
\draw (3.5,1)--(4,0.5);
\draw (4,0.5)--(4.5,0.5);
\draw (4,0.5)--(4,0);
\draw (4.1,0)--(4.1,0.45);
\draw (4.1,0.55)--(4.1,3);
\draw (0.5,3)--(4.1,3);
\draw (4.1,3)--(4.6,3.5);
\node at (3.75,2.25) [above] {$\emptyset$};
\node at (1.5,0.5) [right] {$\emptyset$};
\draw[<->,dotted] (0,0.05)--(0,0.95);
\node at (0,0.5) [left] {$Q_2$};
\draw[<->,dotted] (0,1.05)--(0,1.95);
\node at (0,1.5) [left] {$Q_1$};
\draw[<->,dotted] (0,2.05)--(0,2.95);
\node at (0,2.5) [left] {$Q_2$};
\end{tikzpicture}
\end{center}
\caption{Two web diagrams for the 5d $G_2$ pure gauge theory. }
\label{f:web-G2}
\end{figure}

We can also use the extended topological vertex formalism to the 5d $G_2$ pure gauge theory. Indeed,
there are two different ways in the fivebrane web realizations of the 5d $G_2$ pure gauge theory \cite{G-type,Hayashi:2018lyv}. (We take the limit like \eqref{o-vert-soodd} in which the uncolored leg touches the O5-plane in Figure \ref{f:web-G2}.) Applying the extended topological vertex formalism and using the identity \eqref{M-identity} of the $M$-factor, we can show that the two diagrams provide the same formula for the $G_2$ instanton partition function
\bea\label{G2}
Z_{G_2}^{\textrm{inst}}=&\sum_{{\lambda^{(1)}},{\lambda^{(2)}},{\lambda^{(3)}}}(-Q)^{|{\lambda^{(2)}}|+|{\lambda^{(1)}}|}(-QQ_2^2)^{|{\lambda^{(3)}}|}f_{\lambda^{(1)}} f_{\lambda^{(2)}}^{-1}f_{\lambda^{(3)}}^{-2}\prod_{i=1}^3s_{\lambda^{(i)}}(q^{-\rho})s_{{\lambda^{(i)}}^\vee}(q^{-\rho})\cr
&N_{{\lambda^{(1)}}{\lambda^{(2)}}}(Q_1,q)^{-2}N_{{\lambda^{(1)}}{\lambda^{(3)}}}(Q_1Q_2,q)^{-1}N_{{\lambda^{(2)}}{\lambda^{(3)}}}(Q_2,q)^{-1}N_{\emptyset{\lambda^{(1)}}}(Q_2,q)^{-1}\cr
&N_{{\lambda^{(3)}}^\vee\emptyset}(Q_1Q_2^2,q)^{-1}N_{\emptyset{\lambda^{(2)}}}(Q_1Q_2,q)^{-1} N_{{\lambda^{(3)}}^\vee{\lambda^{(1)}}}(Q_1Q_2^3,q)^{-1} N_{{\lambda^{(3)}}^\vee{\lambda^{(2)}}}(Q_1^2Q_2^3,q)^{-1}\cr
&M_{{\lambda^{(1)}},{\lambda^{(2)}},{\lambda^{(3)}}}(Q_2,Q_1Q_2,Q_1Q_2^2)~.
\eea
This expression is equal to \cite[(2.53)]{Kim:2018gjo} with the change of variables $Q_2=e^{-v_1},Q_1Q_2=e^{-v_2},Q_1Q_2^2=e^{v_3}$ if we ignore the matter and take the unrefined limit there.

\section{Future directions}
The results in this paper will lead to a variety of natural generalizations. More importantly, we expect that many new research directions will be opened up.

\begin{itemize}
  \item \textbf{More general 5d theories:} As briefly mentioned in \S\ref{sec:ADHM}, the partition function with matters in the fundamental representation can be naturally obtained. However, more analysis has to be carried out to obtain closed-form expressions of instanton partition functions with matters in higher rank representations or spinor representations of $\SO(n)$ \cite{Hwang:2014uwa,Zafrir:2015ftn,Kim:2018gjo} as well as quiver gauge theories with gauge groups of type $BCD$.
    \item \textbf{O-vertex for $\Sp(N)$ instantons:} In \S\ref{sec:O}, we have realized $\SO(n)$ instanton partition functions by applying the extended topological vertex to fivebrane webs with an O5${}^-$-plane in string theory. Although an approach to the $\Sp(N)$ instanton partition functions by the topological vertex has been proposed in \cite{Kim-Yagi}, the forms of the partition functions in \eqref{Sp-ADHM-pm} strongly suggest that there is a formalism of the extended topological vertex analogous to \S\ref{sec:O} for fivebrane webs with an O5${}^+$-plane in string theory.
    \item \textbf{Codimension-4 defects and $qq$-characters:} We can also consider instanton partition functions with various defects. In particular, partition functions with codimension-4 defects appear as the $Y$-operator in the $qq$-characters \cite{BPS/CFT,Kim:2016qqs}.
    At the refined level, the $qq$-characters for gauge groups of type $BCD$ have infinitely many terms \cite{Haouzi:2020yxy}. However, they truncate to finite terms at the unrefined level. This allows us to study Lie algebra theoretic relations of $qq$-characters, interpretations as quantum Seiberg-Witten curves, and actions of quantum toroidal algebras, which will be addressed in the forthcoming paper \cite{Nawata-Zhu}.
    \item \textbf{Relation with blowup equations:} The blowup equation, originally introduced in \cite{Gottsche:2006bm,Nakajima:2009qjc}, provides a powerful tool to compute the instanton partition function recursively. Given the input data such as the one-loop and one-instanton partition function, one can fix the full instanton partition function completely from the blowup equation in principle. It is necessary to show that our expressions for $BCD$ gauge theories satisfy the corresponding blowup equations developed in \cite{Keller:2012da,Kim:2019uqw} as a future work. On the other hand, the relation between the blowup equation and the topological string was revealed in \cite{Huang:2017mis}, and it will be very interesting to explore the role of an orientifold in this context.
    \item \textbf{AGT correspondence and $W$-algebras:} The AGT correspondence has been studied exclusively for gauge groups of type $A$ \cite{Alday:2009aq} in the last decade because closed-form expressions of instanton partition functions in terms of Young diagrams were available only for type $A$. The results in this paper need to be examined from the viewpoint of the AGT correspondence and W-algebras \cite{Hollands:2010xa,Hollands:2011zc,Keller:2011ek}.
    \item \textbf{Surface defects and isomonodromic deformations:} The discrete Fourier transformations of the unrefined instanton partition functions can be interpreted as the $\tau$-functions of Painleve-type equations   \cite{Gamayun:2012ma}. This can be formulated in the framework of the blowup equations with surface defects
  \cite{Nekrasov:2020qcq,Jeong:2020uxz}. This program was generalized to arbitrary simple gauge groups in \cite{Bonelli:2021rrg} and the connection with the analytic expression obtained in this paper is to be explored in the future.
  \item \textbf{Geometric engineering of O5-plane:} For gauge groups of type $A$, the geometric engineering \cite{Katz:1996fh,Katz:1997eq} connects a 5d gauge theory to a topological string theory of a toric Calabi-Yau three-fold, and a skeleton of the toric Calabi-Yau three-fold is identified with the corresponding fivebrane web \cite{Leung-Vafa}. In this paper, we show that fivebrane webs with an O5${}^-$-plane can be formulated by the extended topological vertex. However, we miss an interpretation of fivebrane webs with an O5${}^-$-plane in terms of a topological string theory of a Calabi-Yau three-fold. In particular, we expect that the instanton partition function obtained in this paper is a generating function of Gopakumar-Vafa invariants \cite{Gopakumar:1998ki} of a certain Calabi-Yau three-fold. It would be intriguing to find a duality between fivebrane webs with an O5-plane and a Calabi-Yau three-fold.
  \item \textbf{Open topological string with an O5-plane:} The topological vertex was originally developed from the large $N$ duality between the topological string theory and Chern-Simons theory. It is natural to ask whether the extended topological vertex can be understood as the large $N$ limit of $\SO(N)/\Sp(N)$ Chern-Simons theory. It is expected that open topological string theory with the O-vertex gives ingenious perspectives of Kauffman polynomials  \cite{Marino:2009mw}.
  \item \textbf{Refinement:} As elucidated in Appendix \ref{app:ADHM}, JK-residues from the ADHM descriptions of type $BCD$ are \emph{not} classified in the conventional way by Young diagrams at the refined level. Nevertheless, it is desirable to obtain closed-form expressions of fully refined instanton partition functions for arbitrary gauge groups like $\SO(7)$ and $G_2$ gauge groups in \cite{Kim:2018gjo}.
\end{itemize}

\acknowledgments
We are grateful to the authors (Sung-Soo and Futoshi) of \cite{Kim-Yagi} and the authors (Hee-Cheol, Joonho, Seok, Ki-Hong, Jaemo)  of \cite{Kim:2018gjo} for explanations and clarifications of their papers. We would also like to thank Hirotaka Hayashi, Yuji Tachikawa, and especially Futoshi Yagi for their comments on the draft. S.N. would like to thank Yau Center, Tsinghua University, and R.Z. is grateful to Sun Yat-sen Univ., ITP (Chinese Academy of Sciences), Peng Huanwu Center for Fundamental Theory (USTC),  for the warm hospitality where part of the work was carried out and preliminary results were presented. The research of S.N. is supported by the National Science Foundation of China under Grant No.12050410234 and Fudan University Original Project (No. IDH1512092/002).

\appendix
\section{Notations and definitions}\label{app:notations}
In this appendix, we summarize notations and definitions necessary for the paper. For more details, we refer to \cite{Hayashi:2020hhb}.
Let $\lambda=\left(\lambda_1,\lambda_2,\cdots\right)$ be a Young diagram (i.e. non-negative integers such that $\lambda_i \geq \lambda_{i+1}$ and $|\lambda|=\sum_i\lambda_i<\infty$). Let $\ell(\lambda)$ be the length of the Young diagram, which is the number of non-zero $\lambda_i$. We write the transposition of $\lambda$ by $\lambda^\vee$. The arm length $a_{i,j}(\lambda)$ of a box at $(i,j)$ is the number of boxes to the right of the box in the diagram $\lambda$, and the leg length $l_{i,j}(\lambda)$ is the number of boxes below the box. If it is clear from the context, we simply write $a_{i,j}$ and $l_{i,j}$.
\begin{figure}[ht]\centering
  \begin{tikzpicture}[scale=0.8]
\draw (0,0)--(0,-5)--(1,-5)--(1,-4)--(3,-4)--(3,-2)--(5,-2)--(5,-1)--(6,-1)--(6,0)--(0,0);
\draw (1.9,-1.2)--(1.9,-1.8)--(2.5,-1.8)--(2.5,-1.2)--(1.9,-1.2);
\draw [<->,>=stealth] (2.5,-1.5)--(5,-1.5);
\node [scale=0.8] at (2.2,-1.5)  {$i,j$};
\draw [<->,>=stealth] (2.2,-1.8)--(2.2,-4);
\draw (3.8,-1.2) node [scale=1] {$a_{i,j}$};
\draw (1.7,-2.8) node [scale=1] {$l_{i,j}$};
  \end{tikzpicture}
  \caption{Arm and leg length of a box $(i,j)$ in a Young diagram.}
\end{figure}
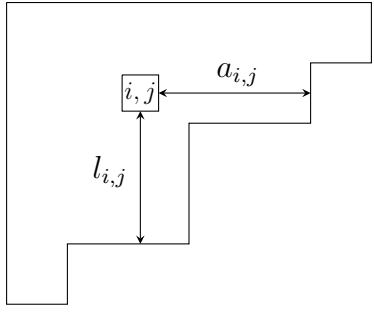

Let us fix the notation of 5d instanton partition functions.
Moduli spaces of instantons receive equivariant actions of $\SO(2)_{\e_1}\times\SO(2)_{\e_1}$ of the space-time and those of $\prod_{s=1}^{{\textrm{rank}G}}\U(1)_{ a_s}$ of the gauge group $G$. Hence, instanton partition functions \cite{Nekrasov:2002qd} depend on the equivariant parameters defined by
$$
q=e^{-\e_1}~, \quad t=e^{\e_2}~,  \quad A_s=e^{- a_s}~.
$$
Note that $A_s$ are called Coulomb branch parameters.
Often we use the notation $2\e_+=\e_1+\e_2$, and it becomes zero in the unrefined limit $\e_1=-\e_2=\hbar$. Therefore, unrefined instanton partition function depends only on $q=e^{-\hbar}$.
In this paper as well as the original $\U(N)$ case \cite{Nekrasov:2002qd}, a $k$-instanton partition function $Z_G^k$ is expressed as a sum over $N$-tuples of Young diagrams, which we denote
$$
\vec{\lambda}=(\lambda^{(1)},\ldots,\lambda^{(N)})~,
$$
where the total number of boxes satisfy
$$
k=\sum_{s=1}^N|\lambda^{(s)}|~.
$$
The total instanton partition function of gauge group $G$ is expressed as a generating function with instanton counting parameter $\frakq$
$$
Z_G=\sum_{k=0}^\infty \frakq^k Z^k_G(A_i;q)~.
$$

Next, let us fix the notation in the formalism of the topological vertex.
We follow the standard notation of the unrefined topological string, in which a Schur function $s_\lambda(x)$ and a skew Schur function $s_{\lambda/\mu}(x)$ play pivotal roles.
The topological string partition function can be constructed from fermionic operators, $\psi$ and $\psi^\ast$, and bosonic current $J$ \cite{IKV,Okounkov:2003sp,Nekrasov:2003rj}
\begin{align}
&\{\psi_n,\psi_m\}=\{\psi_n^\ast,\psi_m^\ast\}=0,\quad \{\psi_n,\psi^\ast_m\}=\delta_{n+m,0},\quad J_n:=\sum_{j\in \mathbb{Z}+1/2}:\psi_{-j}\psi^\ast_{j+n}:~,\cr
&\lt[J_n,\psi_k\rt]=\psi_{n+k},\quad \lt[J_n,\psi^\ast_k\rt]=-\psi^\ast_{n+k},\quad \lt[J_n,J_m\rt]=n\delta_{n+m,0}~.\nonumber
\end{align}
Given the Frobenius coordinate $\lambda=( \alpha_1, \alpha_2,\dots|\beta_1,\beta_2\dots)$ \cite{Macdonald-book}, we define a state
\begin{align}
\ket{\lambda}=(-1)^{\beta_1+\beta_2+\dots+\beta_s+\frac{s}{2}}\psi^\ast_{-\beta_1}\psi^\ast_{-\beta_2}\dots\psi^\ast_{-\beta_s}\psi_{- \alpha_s}\psi_{- \alpha_{(s-1)}}\dots \psi_{- \alpha_1}\ket{0}~.\label{Frobenius-basis}
\end{align}
Then, the skew Schur function can be expressed as
\ba
s_{\lambda/\mu}(\vec{x})=\bra{\mu}V_+(\vec{x})\ket{\lambda}=\bra{\lambda}V_-(\vec{x})\ket{\mu},\label{skew-schur}
\ea
where
\ba\label{VO}
V_\pm (\vec{x})=\exp\lt(\sum_{n=1}^\infty \frac{1}{n}\sum_i x_i^n J_{\pm n}\rt)~.
\ea

The topological vertex labeled by Young diagrams $\nu,\mu,\lambda$ in the clockwise direction
is given by
\be
C_{\nu\mu\lambda}=q^{\frac{\kappa(\mu)}{2}+\frac{\kappa(\lambda)}{2}}s_{\lambda}(q^{-\rho})\sum_{\sigma}s_{\nu^\vee/\sigma}(q^{-\rho-\lambda})s_{\mu/\sigma}(q^{-\rho-\lambda^\vee})~,\label{topvertex}
\ee
where $\kappa(\lambda):=2\sum_{(i,j)\in\lambda}(j-i)$, and
$$
q^{-\rho-\lambda}=\{q^{i-\frac{1}{2}-\lambda_i}\}_{i=1}^\infty=\{q^{\frac{1}{2}-\lambda_1},q^{\frac{3}{2}-\lambda_2},q^{\frac{5}{2}-\lambda_3},\dots\}~.
$$
The framing factor is given by
\be
f_\lambda:=(-1)^{|\lambda|}q^{\frac{\kappa(\lambda)}{2}}~. \label{framing}
\ee
The unrefined Nekrasov factor is defined by
\ba\label{Nekfactor}
N_{\lambda\nu}(Q,q):=\prod_{(i,j)\in\lambda}(1-Q q^{\lambda_i+\nu^\vee_j-i-j+1})\prod_{(i,j)\in\nu}(1-Qq^{-\nu_i-\lambda^\vee_j+i+j-1})~.
\ea
It can be related to the Schur function as
$$
N_{\lambda\lambda}(1,q)^{-1}=(-1)^{|\lambda|}s_\lambda(q^{-\rho})s_{\lambda^\vee}(q^{-\rho})~.
$$
The vertex functions are defined by
\begin{align}
V_{\nu}:=& \sum_{\substack{\mu,\lambda\\2|\mu|+2|\lambda| = |\nu|}}(-1)^{|\mu|}f_\lambda^{2}C_{\nu\mu\lambda}C_{\emptyset\mu^\vee\lambda}~,  \cr
W_\nu :=&\sum_{\substack{\mu,\lambda\\2|\mu| + 2|\lambda| = |\nu|}}(-1)^{|\mu|}f_\lambda^{-4}C_{\mu\nu\lambda}C_{\mu^\vee\emptyset\lambda}~.\label{Wnu}
\end{align}
A plethystic exponential function is defined by
$$
\textrm{PE}\lt(f(x_1,x_2,\dots,x_n)\rt):=\exp\lt(\sum_{k=1}^\infty \frac{1}{k}f(x_1^k,x_2^k,\dots,x_n^k)\rt)~.
$$

\section{Contour integrals from ADHM}\label{app:ADHM}
\Yboxdim6pt

This appendix analyzes the residue structure of the Jeffrey-Kirwan (JK) contour integrals \cite{jeffrey1995localization,Benini:2013xpa} of instanton partition functions for gauge groups of type $BCD$ obtained from the ADHM descriptions. There are many works \cite{Marino:2004cn,Fucito:2004gi,Hollands:2010xa,Hollands:2011zc,Kim:2012qf,Hwang:2014uwa,Nakamura:2014nha,Nakamura:2015zsa} to carry out the contour integrals for $\SO(n)$ gauge group \cite{Nekrasov-Shadchin} and for $\Sp(N)$ gauge group \cite{Kim:2012gu}. As analyzed in these works, the JK residues for gauge groups of type $BCD$ are \emph{not} classified by the conventional way with Young diagrams as in \cite{Nekrasov:2002qd} at the refined level. Here, we will show that the JK poles are indeed classified by ordered sets of Young diagrams at the \emph{unrefined} level. The analysis here leads to closed-form expressions \eqref{SO-ADHM} and \eqref{Sp-ADHM-pm}.

\subsection{Residues for \texorpdfstring{$\SO(n)$}{SO(n)}}
For the sake of concreteness, let us look at the example of the $\SO(4)$ instanton partition function from \eqref{SO-contour}. Choosing the reference vector as $\eta=1$, the JK residues of the 1-instanton  $Z_{\SO(4)}^{k=1}$ are
\begin{align}\label{SO4-1}
\phi_1=\pm a_s-\e_+~, \qquad s=1,2~,
\end{align}
yielding
\begin{align}\nonumber
Z_{\SO(4)}^{k=1}=&\tfrac{(t-q A_{1}^{2})(t^{2}-q^{2} A_{1}^{2}) A_{2}^{2}}{2(-1+q)(-1+t)(A_{1}-A_{2})(q A_{1}-t A_{2})(-1+A_{1} A_{2})(-t+q A_{1} A_{2})}\cr
&+\tfrac{(q-t A_{1}^{2})(q^{2}-t^{2} A_{1}^{2}) A_{2}^{2}}{2(-1+q)(-1+t)(A_{1}-A_{2})(-t A_{1}+q A_{2})(-1+A_{1} A_{2})(q-t A_{1} A_{2})}\cr
&-\tfrac{A_{1}^{2}(t-q A_{2}^{2})(t^{2}-q^{2} A_{2}^{2})}{2(-1+q)(-1+t)(A_{1}-A_{2})(-t A_{1}+q A_{2})(-1+A_{1} A_{2})(-t+q A_{1} A_{2})}\cr
&-\tfrac{A_{1}^{2}(q-t A_{2}^{2})(q^{2}-t^{2} A_{2}^{2})}{2(-1+q)(-1+t)(A_{1}-A_{2})(q A_{1}-t A_{2})(-1+A_{1} A_{2})(q-t A_{1} A_{2})}~.
\end{align}
When we take the unrefined limit $\e_+=0$, the two residues at $\phi_1=\pm a_{s}$ coincide.
Therefore, the poles essentially correspond to 2-tuples of Young diagrams: $(\yng(1),\emptyset)$ for $\phi_1=\pm a_{1}$  and $(\emptyset,\yng(1))$ for $\phi_1=\pm a_{2}$.

Let us move to the case of 2-instanton with $\eta=(1,1+\varepsilon)$ for a sufficiently small $\varepsilon$. Then, the charge vectors are
\begin{align}\label{2-instanton-vectors}
\mathrm{(i)}.&\ \{e_1,-e_1+e_2\} & \mathrm{(iv)}.&\ \{e_1,e_1+e_2\}\cr
\mathrm{(ii)}.&\ \{-e_2,e_1+e_2\} & \mathrm{(v)}.&\ \{e_2,e_1\}\cr
\mathrm{(iii)}.&\ \{e_2,e_1-e_2\}  & \mathrm{(vi)}.&\  \{e_1-e_2,e_1+e_2\}
\end{align}
The corresponding JK poles for $\{\phi_1,\phi_2\}$ are as follows:
\begin{align}\label{2-instanton-poles}
\mathrm{(i)}.&\ \{\pm a_s-\e_+,\pm a_s-\e_+-\e_j\} \cr
\mathrm{(ii)}.&\ \{\pm a_s-\e_+-\e_j,\mp a_s+\e_+\} \cr
\mathrm{(iii)}.&\ \{\pm a_s-\e_+-\e_j,\pm a_s-\e_+\} \\
\mathrm{(iv)}.&\  \{\pm a_s-\e_+,\mp a_s-\e_++\e_j\} \cr
\mathrm{(v)}.&\ \{\pm a_s-\e_+,\pm a_{s+1}-\e_+\}\cup \{\pm a_s-\e_+,\mp a_{s+1}-\e_+\}\cr
\mathrm{(vi)}.&\ \{-\epsilon_{j},  0\}\cup\{-\e_++\e_j,-\e_+\}\cup\{-\epsilon_{j}+\pi i,\pi i\}\cup\{ -\e_++\e_j+\pi i,-\e_++\pi i\}\nonumber
\end{align}
where $s=1,2$ and $j=1,2$. Hence, even if we take into account the difference by the sign in front of the Coulomb branch parameter, the poles in the forth do not appear in the computation of $\U(N)$ instanton partition functions, and they do not fit into the conventional pole classification by Young diagrams as in type $A$ at the refined level.

However, once we take the unrefined limit $\e_+=0$, the poles in (i)--(iv) are related by relative signs and the exchange $\phi_1\leftrightarrow\phi_2$. More remarkably, they provide the same residues corresponding to $(\yng(2),\emptyset),(\yng(1,1),\emptyset),(\emptyset,\yng(2)),(\emptyset,\yng(1,1))$ in \eqref{SO-ADHM}. Also, all the poles in (v) give the same residue corresponding to $(\yng(1),\yng(1))$. All the residues from (vi) are zero. Consequently, we obtain the unrefined instanton partition function in terms of a sum over 2-tuples of Young diagrams
\begin{align}\nonumber
Z_{\SO(4)}^{k=2}=&\tfrac{q^{2}\left(q-A_{1}^{2}\right)^{2}\left(q^{2}-A_{1}^{2}\right)^{2} A_{2}^{4}}{(-1+q)^{4}(1+q)^{2}\left(A_{1}-A_{2}\right)^{2}\left(A_{1}-q A_{2}\right)^{2}\left(q-A_{1} A_{2}\right)^{2}\left(-1+A_{1} A_{2}\right)^{2}}\cr
&+\tfrac{q^{2}\left(-1+q A_{1}^{2}\right)^{2}\left(-1+q^{2} A_{1}^{2}\right)^{2} A_{2}^{4}}{(-1+q)^{4}(1+q)^{2}\left(A_{1}-A_{2}\right)^{2}\left(-q A_{1}+A_{2}\right)^{2}\left(-1+A_{1} A_{2}\right)^{2}\left(-1+q A_{1} A_{2}\right)^{2}}\cr
&+\tfrac{q^{2} A_{1}^{4}\left(q-A_{2}^{2}\right)^{2}\left(q^{2}-A_{2}^{2}\right)^{2}}{(-1+q)^{4}(1+q)^{2}\left(A_{1}-A_{2}\right)^{2}\left(-q A_{1}+A_{2}\right)^{2}\left(q-A_{1} A_{2}\right)^{2}\left(-1+A_{1} A_{2}\right)^{2}}\cr
&+\tfrac{q^{2} A_{1}^{4}\left(-1+q A_{2}^{2}\right)^{2}\left(-1+q^{2} A_{2}^{2}\right)^{2}}{(-1+q)^{4}(1+q)^{2}\left(A_{1}-A_{2}\right)^{2}\left(A_{1}-q A_{2}\right)^{2}\left(-1+A_{1} A_{2}\right)^{2}\left(-1+q A_{1} A_{2}\right)^{2}}\cr
&+\tfrac{q^{6} A_{1}^{2}\left(-1+A_{1}^{2}\right)^{2} A_{2}^{2}\left(-1+A_{2}^{2}\right)^{2}}{(-1+q)^{4}\left(-q A_{1}+A_{2}\right)^{2}\left(A_{1}-q A_{2}\right)^{2}\left(q-A_{1} A_{2}\right)^{2}\left(-1+q A_{1} A_{2}\right)^{2}}~,
\end{align}
where the terms correspond to Young diagrams $(\yng(2),\emptyset),(\yng(1,1),\emptyset),(\emptyset,\yng(2)),(\emptyset,\yng(1,1)),(\yng(1),\yng(1))$, respectively.

It is also worth mentioning about the shift by $\pi i$ in (vi) of \eqref{2-instanton-poles}, called the \emph{holonomy saddle} \cite{Hwang:2017nop,Hwang:2018riu}. This shift appears because the determinant of the corresponding charge vectors in \eqref{2-instanton-vectors} is greater than one, but it is indeed equal to two in this case. Even at the refined level, the $\SO(4)$ instanton partition function receives no contribution from the poles in (vi) of \eqref{2-instanton-poles}. On the other hand, the $\SO(5)$ instanton partition function has the same pole structure as in \eqref{2-instanton-poles}, and it receives a \emph{non-trivial} contribution from the holonomy saddles:
\begin{align}\label{holonomy-saddle}
  &\textrm{R.H.S. of } \eqref{SO-contour}\big|_{n=5,k=2\textrm{ @ poles of (vi)}}=\\
&\tfrac{-q(q-t)^{2}(q^{2}-t) t(q+t)(q-t^{2}) A_{1}^{4} A_{2}^{4}}{4(q-A_{1})(1-A_{1})^{2}(t-A_{1})(1-q A_{1})(t-q A_{1})(q-t A_{1})(1-t A_{1})(q-A_{2})(t-A_{2})(1-A_{2})^{2}(1-q A_{2})(t-q A_{2})(q-t A_{2})(1-t A_{2})}\nonumber
\end{align}
This is the universal feature of $\SO(2N+1)$ instanton partition functions at the refined level, and these residues apparently do not fit into the conventional pole classification by Young diagrams as in type $A$. Nevertheless, as can be seen in \eqref{holonomy-saddle}, it turns out that the contributions from the holonomy saddles vanish at the unrefined level $t=q$.

This structure holds even at higher instantons. Despite the complicated pole structure of the contour integral \eqref{SO-contour} at the refined level illustrated above, the non-trivial JK poles of the $k$-instanton partition function are classified by $\lfloor \frac{n}2 \rfloor$-tuples of Young diagrams with the total number of boxes $k$ as \eqref{phi(s)} \cite{Marino:2004cn,Fucito:2004gi}, and the JK residues yield \eqref{SO-ADHM}.

\subsection{Residues for \texorpdfstring{$\Sp(N)$}{Sp(N)}}
As briefly described in \S\ref{sec:Sp-ADHM}, the ADHM description of the $k$-instanton moduli space leads to supersymmetric quantum mechanics with $\OO(k)$ gauge group so that the instanton partition functions receive contributions $Z^k_\pm$ from each connected component of $\OO(k)$. The complete contour integral formulas are given in \cite{Kim:2012gu,Kim:2012qf,Hwang:2014uwa} based on the result in \cite{Nekrasov-Shadchin}, and we summarize the formulas below.
Writing
$k=2\ell+\chi$, with $\chi=0,1$, $Z_\pm^k$ are given by
$$
  Z^k_\pm=\frac{1}{|W|}\oint[d\phi]Z_{\textrm{vec},\pm}^k\ .
$$
The order $|W|$ of the Weyl group for $\OO(k)_{\pm}$ is
$$
    \hspace{-1cm}|W|_{+}^{\chi=0} = \frac{1}{2^{\ell-1} \ell!} ,\ |W|_{+}^{\chi=1} = \frac{1}{2^\ell \ell!} ,\ |W|^{\chi=0}_{-} =  \frac{1}{2^{\ell-1}(\ell-1)!},\ |W|^{\chi=1}_{-} = \frac{1}{2^\ell \ell!}.
$$
The integrands for the 5d vector multiplet are given as follows:
\begin{align}
  \hspace*{-0.5cm}Z_{\textrm{vec},+}^{k=2\ell+\chi}=&
	    \left(\frac{1}{2\sinh{ \frac{\e_{1,2}}{2}} \, \prod\limits_{i=1}^{N} 2\sinh{ \frac{\pm  a_i + \epsilon_+}{2}}} \cdot \prod_{I=1}^{\ell} \frac{ 2\sinh{\tfrac{\pm \phi_I}{2}} 2\sinh{ \frac{2\epsilon_+\pm \phi_I }{2}}} {2\sinh{ \frac{\e_{1,2}\pm \phi_I }{2}}}\right)^{\chi}
\cr
	    &\times \prod_{I=1}^{\ell} \frac{2 \sinh{\epsilon_+} }{ 2 \sinh{\frac{\e_{1,2} }{2}}  2\sinh{ \frac{\e_{1,2}\pm 2\phi_{I}}{2}} \, \prod\limits_{i=1}^{N} 2\sinh{ \frac{\epsilon_+\pm \phi_{I} \pm  a_i}{2}} }
		\prod_{I < J}^{\ell} \frac{ 2\sinh{ \tfrac{ \pm \phi_I \pm \phi_J}{2}} 2\sinh{ \frac{2\epsilon_+\pm \phi_{I} \pm \phi_{J}}{2} }}{2\sinh{ \frac{\e_{1,2}\pm \phi_{I} \pm \phi_{J} }{2}}}\cr
    \hspace*{-0.5cm}Z_{\textrm{vec},-}^{k = 2\ell}=& \frac{2\cosh{\epsilon_+}}{2\sinh{ \frac{\e_{1,2}}{2}} \,2\sinh{ \e_{1,2}} \, \prod\limits_{i=1}^{N} 2\sinh{ (\epsilon_+\pm  a_i)}} \cdot \prod\limits_{I=1}^{\ell-1} \frac{ 2\sinh{(\pm \phi_I)} 2\sinh{ (2\epsilon_+\pm \phi_I) } } {2\sinh{ (\e_{1,2}\pm \phi_I)} } \nonumber \\
    	    &\times \prod_{I=1}^{\ell-1} \frac{2 \sinh{\epsilon_+} }{ 2 \sinh{\frac{\e_{1,2} }{2}}  2\sinh{ \frac{\e_{1,2}\pm 2\phi_{I}}{2}} \, \prod\limits_{i=1}^{N} 2\sinh{ \frac{\epsilon_+\pm \phi_{I} \pm  a_i}{2}} }
    		\prod_{I < J}^{\ell-1} \frac{ 2\sinh{ \tfrac{ \pm \phi_I \pm \phi_J}{2}} 2\sinh{ \frac{2\epsilon_+\pm \phi_{I} \pm \phi_{J}}{2} }}{2\sinh{ \frac{\e_{1,2}\pm \phi_{I} \pm \phi_{J} }{2}}} \cr
\hspace*{-0.5cm}Z_{\textrm{vec},-}^{k=2\ell+1}=& \frac{1}{2\sinh{ \frac{\e_{1,2}}{2}} \, \prod\limits_{i=1}^{N} 2\cosh{ \frac{\epsilon_+\pm  a_i}{2}}} \cdot \prod_{I=1}^{\ell} \frac{ 2\cosh{\tfrac{\pm \phi_I}{2}} 2\cosh{ \frac{ 2\epsilon_+\pm \phi_I }{2}}} {2\cosh{ \frac{\e_{1,2}\pm \phi_I}{2}}}\cr
	    &\times \prod_{I=1}^{\ell} \frac{2 \sinh{\epsilon_+} }{ 2 \sinh{\frac{\e_{1,2} }{2}}  2\sinh{ \frac{\e_{1,2}\pm 2\phi_{I}}{2}} \, \prod\limits_{i=1}^{N} 2\sinh{ \frac{\epsilon_+\pm \phi_{I} \pm  a_i }{2}} }
		\prod_{I < J}^{\ell} \frac{2\sinh{ \tfrac{ \pm \phi_I \pm \phi_J}{2}} 2\sinh{ \frac{2\epsilon_+\pm \phi_{I} \pm \phi_{J}}{2} }}{2\sinh{ \frac{\e_{1,2}\pm \phi_{I} \pm \phi_{J}}{2}}}~.\label{Sp-contour}
	\end{align}

Now let us analyze the JK residues of these integrals.
Since the JK poles at the refined level are rather complicated, we focus only on the unrefined limit with $\Sp(1)$ gauge group. We will see below that the JK residues are expressed as a sum over 5-tuples of Young diagrams as in \eqref{Sp-ADHM-pm}. The integral is trivial at 1-instanton so that we will investigate JK residues from 2-instanton.

\subsubsection*{2-instanton}
The JK poles for the integrand $Z_{\textrm{vec},+}^{k=2}$ with $\eta=(+1)$ are located at
\be\label{Sp1-2instanton}
\phi_1=\pm a_1~,\quad\pm\frac{\hbar}2~,\quad\pm\frac{\hbar}2+\pi i~,
\ee
where the first comes from $\sinh(\frac{\phi_1\pm a_1}2)$, and the second and third originate from $\sinh(\frac{2\phi_1\pm\hbar}2)$ in the denominator of $Z_{\textrm{vec},+}^{k=2}$. Again, the charge vectors with determinant two yield the holonomy saddle here. Then,
the corresponding JK residues are given by
\begin{align}
Z_+^{k=2}=&\tfrac{q^{3} A_{1}^{6}}{(-1+q)^{2}\left(\sqrt{q}-A_{1}\right)^{2}\left(-1+A_{1}\right)^{2}\left(1+A_{1}\right)^{2}\left(\sqrt{q}+A_{1}\right)^{2}\left(-1+\sqrt{q} A_{1}\right)^{2}\left(1+\sqrt{q} A_{1}\right)^{2}}\cr
&+\tfrac{q^{4} A_{1}^{2}}{2(-1+\sqrt{q})^{4}(1+\sqrt{q})^{4}(1+q)^{2}\left(\sqrt{q}-A_{1}\right)^{2}\left(-1+\sqrt{q} A_{1}\right)^{2}}\cr
&+\tfrac{q^{4} A_{1}^{2}}{2(-1+\sqrt{q})^{4}(1+\sqrt{q})^{4}(1+q)^{2}\left(\sqrt{q}+A_{1}\right)^{2}\left(1+\sqrt{q} A_{1}\right)^{2}}~.\nonumber
\end{align}
These terms correspond to the contributions from $(\yng(1),\emptyset,\emptyset,\emptyset,\emptyset)$, $(\emptyset,\yng(1),\emptyset,\emptyset,\emptyset)$, $(\emptyset,\emptyset,\yng(1),\emptyset,\emptyset)$, respectively, in \eqref{Sp-ADHM-pm}. It is consistent because the contributions from $(\emptyset,\emptyset,\emptyset,\yng(1),\emptyset)$, $(\emptyset,\emptyset,\emptyset,\emptyset,\yng(1))$ are zero in \eqref{Sp-ADHM-pm}.

\subsubsection*{3-instanton}
The JK poles for the integrand $Z_{\textrm{vec},+}^{k=3}$ with $\eta=(+1)$ are located at
\be\label{Sp1-3instanton-p}
\phi_1=\pm a_1~,\quad\pm\frac{\hbar}2~,\quad\pm\frac{\hbar}2+\pi i~,\quad\pm\hbar~.
\ee
The difference from \eqref{Sp1-2instanton} lies at the last one coming from $\sinh(\frac{\phi_1\pm\hbar}{2})$. Then, the corresponding JK residues are given by
$$
\begin{aligned}
Z_+^{k=3}=&\tfrac{q^{6} A_{1}^{7}}{2(-1+q)^{4}\left(\sqrt{q}-A_{1}\right)^{2}\left(q-A_{1}\right)^{2}\left(1+A_{1}\right)^{2}\left(\sqrt{q}+A_{1}\right)^{2}\left(-1+\sqrt{q} A_{1}\right)^{2}\left(1+\sqrt{q} A_{1}\right)^{2}\left(-1+q A_{1}\right)^{2}}\cr
&+\tfrac{q^{6} A_{1}^{3}}{4(-1+\sqrt{q})^{6}(1+\sqrt{q})^{6}(1+q)^{2}(1+\sqrt{q}+q)^{2}\left(\sqrt{q}-A_{1}\right)^{2}\left(-1+A_{1}\right)^{2}\left(-1+\sqrt{q} A_{1}\right)^{2}}\cr
& +\tfrac{q^{6} A_{1}^{3}}{4(-1+\sqrt{q})^{6}(1+\sqrt{q})^{6}(1+q)^{2}(1-\sqrt{q}+q)^{2}\left(-1+A_{1}\right)^{2}\left(\sqrt{q}+A_{1}\right)^{2}\left(1+\sqrt{q} A_{1}\right)^{2}} \cr
&+ \tfrac{q^{B} A_{1}^{3}}{2(-1+\sqrt{q})^{6}(1+\sqrt{q})^{6}(1+q)^{2}(1-\sqrt{q}+q)^{2}(1+\sqrt{q}+q)^{2}\left(q-A_{1}\right)^{2}\left(-1+A_{1}\right)^{2}\left(-1+q A_{1}\right)^{2}}~.
\end{aligned}
$$
These terms correspond to the contributions from $(\yng(1),\emptyset,\emptyset,\emptyset,\emptyset)$, $(\emptyset,\yng(1),\emptyset,\emptyset,\emptyset)$, $(\emptyset,\emptyset,\yng(1),\emptyset,\emptyset)$, $(\emptyset,\emptyset,\emptyset,\yng(1),\emptyset)$, respectively, in \eqref{Sp-ADHM-pm}. It is consistent because the contribution from $(\emptyset,\emptyset,\emptyset,\emptyset,\yng(1))$ is zero in \eqref{Sp-ADHM-pm}.

The JK poles for the integrand $Z_{\textrm{vec},-}^{k=3}$ with $\eta=(+1)$ are located at
\be\label{Sp1-3instanton-m}
\phi_1=\pm a_1~,\quad\pm\frac{\hbar}2~,\quad\pm\frac{\hbar}2+\pi i~,\quad\pm\hbar+\pi i~.
\ee
The difference from \eqref{Sp1-3instanton-p} lies at the last one coming from $\cosh(\frac{\phi_1\pm\hbar}{2})$. Then, the corresponding JK residues are given by
$$
\begin{aligned}
Z_-^{k=3}=-& \tfrac{q^{6} A_{1}^{7}}{2(-1+q)^{4}\left(\sqrt{q}-A_{1}\right)^{2}\left(-1+A_{1}\right)^{2}\left(\sqrt{q}+A_{1}\right)^{2}\left(q+A_{1}\right)^{2}\left(-1+\sqrt{q} A_{1}\right)^{2}\left(1+\sqrt{q} A_{1}\right)^{2}\left(1+q A_{1}\right)^{2}}\\
&-\tfrac{q^{6} A_{1}^{3}}{4(-1+q)^{6}(1+q)^{2}(1-\sqrt{q}+q)^{2}\left(\sqrt{q}-A_{1}\right)^{2}\left(1+A_{1}\right)^{2}\left(-1+\sqrt{q} A_{1}\right)^{2}}-\\
&- \tfrac{q^{6} A_{1}^{3}}{4(-1+q)^{6}(1+q)^{2}(1+\sqrt{q}+q)^{2}\left(1+A_{1}\right)^{2}\left(\sqrt{q}+A_{1}\right)^{2}\left(1+\sqrt{q} A_{1}\right)^{2}}-\\
& -\tfrac{q^{8} A_{1}^{3}}{2(-1+q)^{6}(1+q)^{2}(1-\sqrt{q}+q)^{2}(1+\sqrt{q}+q)^{2}\left(1+A_{1}\right)^{2}\left(q+A_{1}\right)^{2}\left(1+q A_{1}\right)^{2}}~.
\end{aligned}
$$
These terms correspond to the contributions from $(\yng(1),\emptyset,\emptyset,\emptyset,\emptyset)$, $(\emptyset,\yng(1),\emptyset,\emptyset,\emptyset)$, $(\emptyset,\emptyset,\yng(1),\emptyset,\emptyset)$, $(\emptyset,\emptyset,\emptyset,\yng(1),\emptyset)$, respectively, in \eqref{Sp-ADHM-pm}. It is consistent because the contribution from $(\emptyset,\emptyset,\emptyset,\emptyset,\yng(1))$ is zero in \eqref{Sp-ADHM-pm}.

\subsubsection*{4-instanton}
JK poles for $Z_{\textrm{vec},+}^{k=4}$ are more complicated because it involves two integration variables $\{\phi_1,\phi_2\}$. With the choice of $\eta=(1,1+\varepsilon)$, the charge vectors with determinant one are as (i)--(v) in \eqref{2-instanton-vectors}.
The corresponding JK poles are
\begin{align}\nonumber
\{\pm a_1,\pm a_1+\hbar \}\cup \{\pm a_1,\mp a_1-\hbar \}\cup\{\pm a_1+\hbar,\pm a_1 \}\cup \{\mp a_1-\hbar ,\pm a_1 \}\cr
\{\pm a_1,\pm a_1-\hbar \}\cup \{\pm a_1,\mp a_1+\hbar \}\cup\{\pm a_1-\hbar,\pm a_1  \}\cup \{\mp a_1+\hbar,\pm a_1  \}~.
\end{align}
The poles at the first line give the term for $(\yng(2),\emptyset,\emptyset,\emptyset,\emptyset)$ in
\eqref{Sp-ADHM-pm} whereas those at the second line provide the term for $(\yng(1,1),\emptyset,\emptyset,\emptyset,\emptyset)$.

The charge vectors with determinant two are
\begin{align}\nonumber
  \mathrm{(a)}.&\ \{2e_1,-e_1+e_2\} & \mathrm{(e)}.&\ \{e_2,2e_1\}\cr
  \mathrm{(b)}.&\ \{-2e_2,e_1+e_2\} & \mathrm{(f)}.&\ \{2e_2,e_1\}\cr
  \mathrm{(c)}.&\ \{2e_2,e_1-e_2\}  & \mathrm{(g)}.&\  \{e_1-e_2,e_1+e_2\}\cr
  \mathrm{(d)}.&\ \{2e_1,e_1+e_2\} & &
\end{align}
The JK poles corresponding to (a)--(d) are located at
\begin{align}\nonumber
\left\{\pm \frac{\hbar}2,\pm \frac{3\hbar}2\right\}\cup\left\{\pm \frac{3\hbar}2,\pm \frac{\hbar}2\right\}\cr
\left\{\pm \frac{\hbar}2,\mp \frac{\hbar}2\right\}\cup\left\{\mp \frac{\hbar}2,\pm \frac{\hbar}2\right\}
\end{align}
as well as their simultaneous $\pi i$ shifts. The poles from the first line give the term for $(\emptyset,\yng(1,1),\emptyset,\emptyset,\emptyset)$ in \eqref{Sp-ADHM-pm} whereas at the second line provide the term for $(\emptyset,\yng(2),\emptyset,\emptyset,\emptyset)$. The terms for  $(\emptyset,\emptyset,\yng(1,1),\emptyset,\emptyset)$ and $(\emptyset,\emptyset,\yng(2),\emptyset,\emptyset)$ come from their holonomy saddles.
The JK poles corresponding to (e),(f) are located at
$$
\left\{\pm \frac{\hbar}2,\pm  a_1 \right\}\cup\left\{\pm \frac{\hbar}2,\mp  a_1 \right\}\cup\left\{\pm  a_1 ,\pm \frac{\hbar}2\right\}\cup\left\{\mp  a_1 ,\pm \frac{\hbar}2\right\}
$$
as well as the $\pi i$ shifts $\pm \frac{\hbar}2\to \pm \frac{\hbar}2+\pi i$. These poles yield the residues corresponding to $(\yng(1),\yng(1),\emptyset,\emptyset,\emptyset)$ and $(\yng(1),\emptyset,\yng(1),\emptyset,\emptyset)$ in \eqref{Sp-ADHM-pm}. The JK poles corresponding to (g) are located at
$$
\left\{0,\pm\hbar \right\}\cup\left\{\pm\hbar ,0\right\}\cup
\left\{0,\pm\hbar +\pi i\right\}\cup\left\{\pm\hbar+\pi i ,0\right\}
$$
These poles yield the terms for $(\emptyset,\emptyset,\emptyset,\yng(2),\emptyset)$, $(\emptyset,\emptyset,\emptyset,\yng(1,1),\emptyset)$, $(\emptyset,\emptyset,\emptyset,\emptyset,\yng(2))$, $(\emptyset,\emptyset,\emptyset,\emptyset,\yng(1,1))$ in \eqref{Sp-ADHM-pm}.

The charge vectors with determinant four consist of $\{2e_2,2e_1\}$. The poles with non-trivial residues are located at
$$
 \left\{\pm\frac\hbar2,\pm\frac\hbar2+\pi i \right\}\cup \left\{\pm\frac\hbar2+\pi i,\pm\frac\hbar2 \right\}~,
$$
and the residues sum up to the term for $(\emptyset,\yng(1),\yng(1),\emptyset,\emptyset)$ in \eqref{Sp-ADHM-pm}.

The contributions from $(\yng(1),\emptyset,\emptyset,\yng(1),\emptyset)$, $(\yng(1),\emptyset,\emptyset,\emptyset,\yng(1))$, $(\emptyset,\yng(1),\emptyset,\yng(1),\emptyset)$, $(\emptyset,\yng(1),\emptyset,\emptyset,\yng(1))$, $(\emptyset,\emptyset,\yng(1),\yng(1),\emptyset)$, $(\emptyset,\emptyset,\yng(1),\emptyset,\yng(1))$, $(\emptyset,\emptyset,\emptyset,\yng(1),\yng(1))$ are trivial in \eqref{Sp-ADHM-pm}. In total, the 4-instanton partition function can be written as a sum over 5-tuples of Young diagrams with the total number of boxes two as in \eqref{Sp-ADHM-pm}.

On the other hand, JK poles for the other part $Z_{\textrm{vec},-}^{k=4}$ are simpler. The JK poles with determinant one are located at $\phi_1=\pm a_1$, which corresponds to $(\yng(1),\emptyset,\emptyset,\emptyset,\emptyset)$.
The JK poles with determinant two are located at
$$
\pm \frac{\hbar}{2}, \quad \pm \frac{\hbar}{2}+\pi i, \quad \pm \hbar, \quad \pm \hbar+\pi i,
$$
corresponding to $(\emptyset,\yng(1),\emptyset,\emptyset,\emptyset)$, $(\emptyset,\emptyset,\yng(1),\emptyset,\emptyset)$, $(\emptyset,\emptyset,\emptyset,\yng(1),\emptyset)$, $(\emptyset,\emptyset,\emptyset,\emptyset,\yng(1))$,  respectively.

\subsubsection*{higher instantons}

At this moment, we familiarize ourselves with the JK residues of the contour integrals \eqref{Sp-contour}.
Similarly, we can analyze the JK residues at higher instantons.
The JK poles of \eqref{Sp-contour} are classified by 5-tuples of Young diagrams $\lambda^{(s)}$  $(s=1, \cdots, 5)$:
$$
\phi_{i, j}(s)= a_{s}+(i-j) \hbar, \quad(i, j) \in \lambda^{(s)}
$$
where the total number of boxes is equal to the number of integration variables in \eqref{Sp-contour}. As seen above, $a_{1}$ is the Coulomb branch parameter of $\Sp(1)$, and the other parameters are
\bea\label{4otherpoles}
& a_2=\frac\hbar2~, \quad  a_3=\frac\hbar2+\pi i~, \quad  a_4=0~, \quad  a_5=\pi i~, & &\textrm{for } Z_{\textrm{vec},+}^{k=2\ell}\cr
& a_2=\frac\hbar2~, \quad  a_3=\frac\hbar2+\pi i~, \quad  a_4=\hbar~, \quad  a_5=\pi i~, && \textrm{for } Z_{\textrm{vec},+}^{k=2\ell+1}\cr
& a_2=\frac\hbar2~, \quad  a_3=\frac\hbar2+\pi i~, \quad  a_4=\hbar~, \quad  a_5=\hbar+\pi i~, && \textrm{for } Z_{\textrm{vec},-}^{k=2\ell}\cr
& a_2=\frac\hbar2~, \quad  a_3=\frac\hbar2+\pi i~, \quad  a_4=0~, \quad  a_5=\hbar+\pi i~, && \textrm{for } Z_{\textrm{vec},-}^{k=2\ell+1}~.\eea
\emph{Up to constants}, it is straightforward though tedious to derive that the residues at these poles are given by \eqref{Sp-ADHM-pm}, which can be fittingly packaged into the single formula \eqref{Sp-ADHM}.

In principle, the constants in \eqref{Sp-ADHM} are determined by the number of poles with the same residue and the determinant of the corresponding charge vectors. By performing explicit calculations at high instanton numbers, we propose the following conjectural expressions. (It would be desirable to derive the expressions rigorously.)
For effective Coulomb branch parameters $A_s=\pm1,\pm q^{\frac12}$, we assign the following coefficient to a Young diagram $\lambda^{(s)}$ that is created by adding boxes to the right and bottom of the $m\times m$ rectangular Young diagram
\be\label{const1}
{\raisebox{-1.5cm}{\includegraphics[width=5cm]{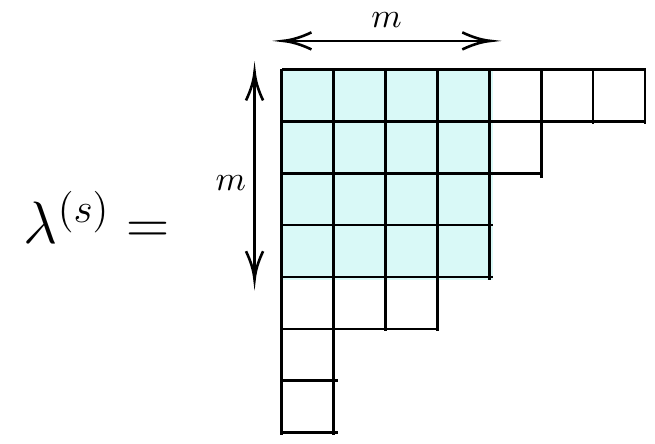}}} \quad \leadsto \quad C_{\lambda^{(s)},A_s=\pm1,\pm q^{\frac12}}=\frac{2^{2m-1}}{\binom{2m-1}{m-1}}~.
\ee
For effective Coulomb branch parameters $A_s=\pm q$, we assign the following coefficient to a Young diagram $\lambda^{(s)}$ that is created by adding boxes to the right and bottom of the $m\times (m+1)$ rectangular Young diagram
\be\label{const2}
{\raisebox{-1.5cm}{\includegraphics[width=5cm]{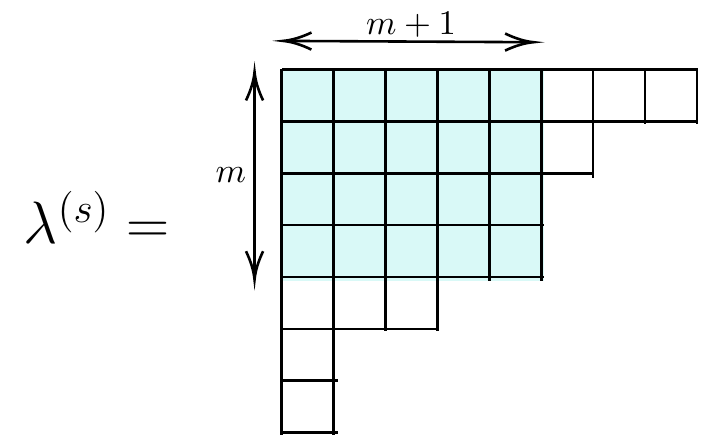}}} \quad \leadsto \quad C_{\lambda^{(s)},A_s=\pm q}=\frac{2^{2m}}{\binom{2m+1}{m}}~.
\ee
Note that a Young diagram with a single column corresponds to $m=0$.

The analysis for higher rank gauge groups goes in parallel. In conclusion, we obtain a sum \eqref{Sp-ADHM-pm} over $(N+4)$-tuples of Young diagrams after the JK-residue integral of \eqref{Sp-contour} for $\Sp(N)$ gauge group.

\section{Derivation of the \texorpdfstring{$M$}{M}-factor}\label{app:M-proof}
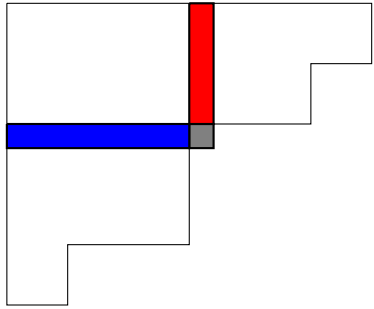
\begin{figure}[ht]\centering
  \begin{tikzpicture}[scale=0.8]
\draw (0,0)--(0,-5)--(1,-5)--(1,-4)--(3,-4)--(3,-2)--(5,-2)--(5,-1)--(6,-1)--(6,0)--(0,0);
\draw[thick,fill=gray] (3,-2)--(3,-2.4)--(3.4,-2.4)--(3.4,-2)--(3,-2);
\draw[thick,fill=red] (3,0)--(3,-2)--(3.4,-2)--(3.4,0)--(3,0);
\draw[thick,fill=blue] (0,-2)--(3,-2)--(3,-2.4)--(0,-2.4)--(0,-2);
  \end{tikzpicture}
  \caption{We add one box at $(m,n)$ colored by gray to $\lambda^{(s)}$ and we call the resulting Young diagram $\widetilde\lambda^{(s)}:=\lambda^{(s)}+\Box$. }\label{fig:Young-onebox}
\end{figure}

In this appendix, we prove the equivalence between \eqref{PE} and \eqref{product} by recursion relation. Let us suppose we add one box at $(m,n)$ to $\lambda^{(s)}$ as in Figure \ref{fig:Young-onebox} and we denote the resulting Young diagram by $\widetilde\lambda^{(s)}:=\lambda^{(s)}+\Box$. Then, it follows from \eqref{PE} that the ratio of the $M$-factors is expressed in the form of a plethystic exponential function
\bea\nonumber
\frac{M_{\lambda^{(1)},\ldots,\lambda^{(s)}+\Box,\ldots,\lambda^{(N)}}}{M_{\lambda^{(1)}\ldots,\lambda^{(s)},\ldots,\lambda^{(N)}}}=\text{PE}(Y)
\eea
where
\bea\nonumber
Y=&-A_{s}^{2}q^{m-1-\lambda_m^{(s)}}\Bigl[(1+q)q^{m-1-\lambda_m^{(s)}}-q^{\ell(\lambda^{(s)})} +(q-1)\sum_{i\neq m}q^{i-1-\lambda_i^{(s)}} \Bigr] \cr
&+A_{s}q^{m-1-\lambda_m^{(s)}}\sum_{t\neq s}A_{t} \Bigl[ q^{\ell(\lambda^{(t)})}+(1-q)\sum_{i=1}^{\ell(\lambda^{(t)})}q^{i-1-\lambda^{(t)}_i}\bigr)\Bigr]\cr
:=& -A_{s}^{2}Y_1 +A_{s}\sum_{t\neq s}A_{t}Y_2^{(t)}~.
\eea
Now, we will first bring the part of $Y_1$ into a product form. We can manipulate $Y_1$ into
\bea\nonumber
Y_1=&q^{m-1-\lambda_m^{(s)}}\Bigl[{\color{red}{(q-1)\sum_{i< m}q^{i-1-\lambda_i^{(s)}} }}+{\color{gray}{q^{m-1-\lambda_m^{(s)}}}}+{\color{blue}{q^{m-\lambda_m^{(s)}}+(q-1)\sum_{i> m}q^{i-1-\lambda_i^{(s)}}-q^{\ell(\lambda^{(s)})}}} \Bigr] \cr
=&\sum_{(i,j)\in \textrm{{\color{red}{red}},{\color{blue}{blue}},{\color{gray}{gray}}}}q^{a_{i,j}(\widetilde\lambda^{(s)})+l_{i,j}(\widetilde\lambda^{(s)})+2(i-\widetilde\lambda_{i}^{(s)})}-\sum_{(i,j)\in \textrm{{\color{red}{red}},{\color{blue}{blue}}}}q^{a_{i,j}(\lambda^{(s)})+l_{i,j}(\lambda^{(s)})+2(i-\lambda_{i}^{(s)})}~.
\eea

The terms with red, gray and blue in the first line correspond to those from boxes in Young diagrams (Figure \ref{fig:Young-onebox}) with the corresponding colors in the second line. Note that the terms with blue provide non-trivial contributions when $\lambda^{(s)}_i\neq \lambda^{(s)}_{i+1}$.
As a result, we have
\bea\label{product-Y1}
\text{PE}(-A_{s}^{2}Y_1)=&\Bigl[\prod_{(i,j)\in \wt\lambda^{(s)}}(1-A_{s}^{2}q^{a_{i,j}(\wt\lambda^{(s)})+l_{i,j}(\wt\lambda^{(s)})+2(i-\wt\lambda_{i}^{(s)})}) \Bigr]\cr
& \Bigl[\prod_{(i,j)\in \lambda^{(s)}}(1-A_{s}^{2}q^{a_{i,j}(\lambda^{(s)})+l_{i,j}(\lambda^{(s)})+2(i-\lambda_{i}^{(s)})})^{-1}\Bigr]~.
\eea

Next, we will manipulate the part of $Y_2^{(t)}$ into a product form by cases. If $\ell(\lambda^{(t)})<n$, we can write it as
\bea\nonumber
Y_2^{(t)}=&q^{m-1-\lambda_m^{(s)}}\Bigl[q^{\ell(\lambda^{(t)})}+(1-q)\sum_{i=1}^{\ell(\lambda^{(t)})}q^{i-1-\lambda^{(t)}_i}\Bigr] \cr
=&\sum_{j=1}^{n}q^{m-1-a_{m,j}(\widetilde\lambda^{(s)})-\lambda^{(t)}_{j}}-\sum_{j=1}^{n-1}q^{m-1-a_{m,j}(\lambda^{(s)})-\lambda^{(t)}_{j}}~,
\eea
which gives
\bea\nonumber
&\text{PE}(A_{s}A_{t}Y_2^{(t)})\\=&\Bigl[\prod_{(i,j)\in\wt\lambda^{(s)}}(1-A_{s} A_{t} q^{i-1-a_{i,j}(\wt\lambda^{(s)})-\lambda^{(t)}_{j}})^{-1}\Bigr]\Bigl[\prod_{(i,j)\in\lambda^{(s)}}(1-A_{s} A_{t} q^{i-1-a_{i,j}(\lambda^{(s)})-\lambda^{(t)}_{j}})\Bigr]~.
\eea
Combining this with \eqref{product-Y1}, we obtain a recursion relation of \eqref{product} for $\lambda^{(s)}\to \wt\lambda^{(s)}$.
If $\ell(\lambda^{(t)})\ge n$, we can write
\bea\nonumber
Y_2^{(t)}=&q^{m-1-\lambda_m^{(s)}} \Bigl[ (1-q)\sum_{i=1}^{n-1}q^{i-1-\lambda^{(t)}_i}+ q^{\ell(\lambda^{(t)})}+(1-q)\sum_{j=n}^{\ell(\lambda^{(t)})}q^{j-1-\lambda^{(t)}_j}\bigr)\Bigr]\cr
=&\sum_{j=1}^{n}q^{m-1-a_{m,j}(\widetilde\lambda^{(s)})-\lambda^{(t)}_{j}}-\sum_{j=1}^{n-1}q^{m-1-a_{m,j}(\lambda^{(s)})-\lambda^{(t)}_{j}}\cr
& \hspace{5cm} + \sum_{j=1}^{\lambda^{(t)}_n} (q^{l_{n,j}(\lambda^{(t)})-j+1+m}-q^{l_{n,j}(\lambda^{(t)})-j+m})~.
\eea
As a result, we have
\bea
&\text{PE}(A_{s}A_{t}Y_2^{(t)})\cr
=&\Bigl[\prod_{(i,j)\in\wt\lambda^{(s)}}(1-A_{s} A_{t} q^{i-1-a_{i,j}(\wt\lambda^{(s)})-\lambda^{(t)}_{j}})^{-1}\Bigr] \Bigl[\prod_{(i,j)\in\lambda^{(t)}}(1-A_{s} A_{t} q^{l_{i,j}(\lambda^{(t)})-j+1+(\wt\lambda^{(s)})^\vee_{i}})^{-1}\Bigr]~\cr
&\Bigl[\prod_{(i,j)\in\lambda^{(s)}}(1-A_{s} A_{t} q^{i-1-a_{i,j}(\lambda^{(s)})-\lambda^{(t)}_{j}})\Bigr] \Bigl[\prod_{(i,j)\in\lambda^{(t)}}(1-A_{s} A_{t} q^{l_{i,j}(\lambda^{(t)})-j+1+(\lambda^{(s)})^\vee_{i}})\Bigr]~.\nonumber
\eea
Combining this with \eqref{product-Y1}, we obtain a recursion relation of \eqref{product} for $\lambda^{(s)}\to \wt\lambda^{(s)}$.

\section{4d instanton partition functions}\label{app:4d}

As a 5d theory is put on $S^1$, the zero radius limit of $S^1$ leads to an effective 4d $\cN=2$ gauge theory. In this sense, it is straightforward to write down 4d instanton partition functions for gauge groups of type $BCD$ from the analytic expressions in this paper.

To take the 4d limit, we need to restore the dependence on $R$, the radius of $S^1$ in the parameters. We replace
$$
 a_s\rightarrow  a_s R~,\quad \hbar\rightarrow \hbar R~,
$$
in (\ref{SO-ADHM}), (\ref{SO-ADHM2}) and (\ref{Sp-ADHM-pm}). As a consequence, all trigonometric functions in the 4d limit $R\rightarrow 0$ reduce to the rational form
$$
\sinh(\frac{Rx}2)\rightarrow Rx~\quad \cosh(\frac{Rx}2)\rightarrow 1~.
$$
We remark that specific power of the radius, $R^{-n_G |\vec{\lambda}|}$, is absorbed into the instanton counting parameter $\frakq$, i.e.
$$
Z_G=\sum_k \frakq^k Z^k_G\rightarrow Z^{\rm 4d}_G=\sum_k \frakq_{4d}^kZ^{4d\ k}_G,\qquad \textrm{as } R\rightarrow 0,
$$
where we denote $\frakq_{4d}:=\frakq R^{-n_G}$ with
$$
n_{\SO(n)}=2n-4~,\qquad
n_{\Sp(N)}=2N+2~.
$$

Taking this limit, we obtain the 4d $k$-instanton partition function for $\SO(n)$ gauge group
\begin{align}\label{SO-4d-ADHM2}
  &Z^{4d\ k}_{\SO(n)}=\cr= &\sum_{\vec{\lambda}}\prod_{s=1}^N\prod_{(i,j)\in\lambda^{(s)}} \tfrac{\lt(2 a_s+\hbar (i-j+(\lambda^\vee)_{j}^{(s)}-\lambda_{i}^{(s)})\rt)^2}{\lt(\phi_{i,j}(s)\rt)^{2\chi}\prod\limits_{t=1}^N N^2_{i,j}(s,t)} \\
  &\prod_{1\le s< t\le N}\tfrac{1}{\prod\limits_{(i,j)\in\lambda^{(s)}}\lt( a_s+ a_t+\hbar(i+j-1-\lambda^{(s)}_{i}-\lambda^{(t)}_{j})\rt)^2\prod\limits_{(m,n)\in\lambda^{(t)}}\lt( a_s+ a_t+\hbar (1-m-n+(\lambda^{(t)})^\vee_{n}+(\lambda^{(s)})^\vee_{m})\rt)^2}~,\nonumber
 \end{align}
where the summation is taken over $N$-tuples of Young diagrams with $k=\sum_{s=1}^N|\lambda^{(s)}|$. The other functions such as $\phi_{i,j}(s)$ and $N_{i,j}(s,t)$ are as in \eqref{phi(s)} and \eqref{Nij}.

Although there are $N+4$ effective Coulomb branch parameters in \eqref{Sp-ADHM-pm}, two of them disappear in  the 4d limit since two parameters have an imaginary part $i\pi$ in \eqref{4otherpoles}, which changes $\sinh$ to $\cosh$. Therefore, the basic building block for 4d $\Sp(N)$ instanton partition functions becomes
 \begin{align} \label{Sp-4d-ADHM}
  &\wt Z_{\Sp(N)}^{4d\ \ell}(a_1,\ldots, a_{N+2};\hbar)
  =\cr= &\sum_{\vec{\lambda}}C_{\vec{\lambda},\vec{a}}\prod_{s=1}^N\prod_{(i,j)\in\lambda^{(s)}} \tfrac{\lt(2 a_s+\hbar (i-j+(\lambda^\vee)_{j}^{(s)}-\lambda_{i}^{(s)})\rt)^2}{\lt(\phi_{i,j}(s)\rt)^{2\chi}\prod\limits_{t=1}^N N^2_{i,j}(s,t)} \\
  &\prod_{1\le s< t\le N}\tfrac{1}{\prod\limits_{(i,j)\in\lambda^{(s)}}\lt( a_s+ a_t+\hbar(i+j-1-\lambda^{(s)}_{i}-\lambda^{(t)}_{j})\rt)^2\prod\limits_{(m,n)\in\lambda^{(t)}}\lt( a_s+ a_t+\hbar (1-m-n+(\lambda^{(t)})^\vee_{n}+(\lambda^{(s)})^\vee_{m})\rt)^2}~,\nonumber
 \end{align}
 where the summation is taken over $(N+2)$-tuples of Young diagrams with $\ell=\sum_{s=1}^N|\lambda^{(s)}|$. Here, $C_{\vec{\lambda},\vec{a}}$ is defined as an appropriate rational version of \eqref{weight}:
 $$
 C_{\vec{\lambda},\vec{a}}=C_{\lambda^{(N+1)},a_{N+1}}C_{\lambda^{(N+2  )},a_{N+2}}~.
$$
where
 \begin{align}\nonumber
 C_{\lambda^{(s)},a_s=0,\frac{\hbar}2}&=\frac{2^{2m-1}}{\binom{2m-1}{m-1}}\qquad  \textrm{  where $m$ is the number of rows with $\lambda_j^{(s)}\ge j$}~,\cr
 C_{\lambda^{(s)},a_s=\hbar}&=\frac{2^{2m}}{\binom{2m+1}{m}}\qquad  \textrm{  where $m$ is the number of rows with $\lambda_j^{(s)}\ge j+1$}~.
 \end{align}
Using this rational function, the 4d $\Sp(N)$ instanton partition functions are expressed as
 \begin{align}\label{Sp-4d-ADHM-pm}
 Z^{4d\ k=2\ell}_+=& 2 \wt Z_{\Sp(N)}^{4d\ \ell}( a_1,\ldots, a_N, a_{N+1}=\frac{\hbar}{2}, a_{N+2}=0;\hbar) \cr
  Z^{4d\ k=2\ell+1}_+=& \frac{1}{\hbar^2 \prod\limits_{i=1}^{N} a^2_{i}} \wt Z_{\Sp(N)}^{4d\ \ell}( a_1,\ldots, a_N, a_{N+1}=\frac{\hbar}{2}, a_{N+2}=\hbar;\hbar)  ~.
 \end{align}
Note that the contribution from $Z^k_-$ vanishes in the 4d limit as it contains positive power of $R$
\bea
Z^{4d\ k=2\ell}_-=& 2(-1)^{N}\frac{R^{2N}}{4\hbar^4\prod\limits_{i=1}^{N} 4 a^2_{i}} \wt Z_{\Sp(N)}^{4d\ \ell-1}( a_1,\ldots, a_N, a_{N+1}=\frac{\hbar}{2},a_{N+2}=\hbar;\hbar) \cr
Z^{4d\ k=2\ell+1}_-=& (-1)^{N}\frac{R^{2N}}{\hbar^2 } \wt Z_{\Sp(N)}^{4d\ \ell}( a_1,\ldots, a_N, a_{N+1}=\frac{\hbar}{2}, a_{N+2}=0;\hbar)~.\nonumber
\eea
This is consistent with the fact that the non-trivial discrete $\theta$-angle exists only in 5d theories.

\bibliographystyle{JHEP}
\bibliography{O-vert}

\providecommand{\href}[2]{#2}\begingroup\raggedright\begin{thebibliography}{10}

\bibitem{Nekrasov:2002qd}
N.A.~Nekrasov, \emph{{Seiberg-Witten prepotential from instanton counting}},
  \href{https://doi.org/10.4310/ATMP.2003.v7.n5.a4}{\emph{Adv. Theor. Math.
  Phys.} {\bfseries 7} (2003) 831}
  [\href{https://arxiv.org/abs/hep-th/0206161}{{\ttfamily hep-th/0206161}}].

\bibitem{Seiberg:1994rs}
N.~Seiberg and E.~Witten, \emph{{Electric - magnetic duality, monopole
  condensation, and confinement in N=2 supersymmetric Yang-Mills theory}},
  \href{https://doi.org/10.1016/0550-3213(94)90124-4}{\emph{Nucl. Phys. B}
  {\bfseries 426} (1994) 19}
  [\href{https://arxiv.org/abs/hep-th/9407087}{{\ttfamily hep-th/9407087}}].

\bibitem{Seiberg:1994aj}
N.~Seiberg and E.~Witten, \emph{{Monopoles, duality and chiral symmetry
  breaking in N=2 supersymmetric QCD}},
  \href{https://doi.org/10.1016/0550-3213(94)90214-3}{\emph{Nucl. Phys. B}
  {\bfseries 431} (1994) 484}
  [\href{https://arxiv.org/abs/hep-th/9408099}{{\ttfamily hep-th/9408099}}].

\bibitem{Pestun:2007rz}
V.~Pestun, \emph{{Localization of gauge theory on a four-sphere and
  supersymmetric Wilson loops}},
  \href{https://doi.org/10.1007/s00220-012-1485-0}{\emph{Commun. Math. Phys.}
  {\bfseries 313} (2012) 71} [\href{https://arxiv.org/abs/0712.2824}{{\ttfamily
  0712.2824}}].

\bibitem{Alday:2009aq}
L.F.~Alday, D.~Gaiotto and Y.~Tachikawa, \emph{{Liouville Correlation Functions
  from Four-dimensional Gauge Theories}},
  \href{https://doi.org/10.1007/s11005-010-0369-5}{\emph{Lett.Math.Phys.}
  {\bfseries 91} (2010) 167} [\href{https://arxiv.org/abs/0906.3219}{{\ttfamily
  0906.3219}}].

\bibitem{Nekrasov:2012xe}
N.~Nekrasov and V.~Pestun, \emph{{Seiberg-Witten geometry of four dimensional
  N=2 quiver gauge theories}},
  \href{https://arxiv.org/abs/1211.2240}{{\ttfamily 1211.2240}}.

\bibitem{NPS}
N.~Nekrasov, V.~Pestun and S.~Shatashvili, \emph{{Quantum geometry and quiver
  gauge theories}},
  \href{https://doi.org/10.1007/s00220-017-3071-y}{\emph{Commun. Math. Phys.}
  {\bfseries 357} (2018) 519}
  [\href{https://arxiv.org/abs/1312.6689}{{\ttfamily 1312.6689}}].

\bibitem{BPS/CFT}
N.~Nekrasov, \emph{{BPS/CFT correspondence: non-perturbative Dyson--Schwinger
  equations and qq-characters}},
  \href{https://doi.org/10.1007/JHEP03(2016)181}{\emph{JHEP} {\bfseries 03}
  (2016) 181} [\href{https://arxiv.org/abs/1512.05388}{{\ttfamily
  1512.05388}}].

\bibitem{Kimura-Pestun}
T.~Kimura and V.~Pestun, \emph{{Quiver W-algebras}},
  \href{https://doi.org/10.1007/s11005-018-1072-1}{\emph{Lett. Math. Phys.}
  {\bfseries 108} (2018) 1351}
  [\href{https://arxiv.org/abs/1512.08533}{{\ttfamily 1512.08533}}].

\bibitem{Braverman:2004vv}
A.~Braverman, \emph{{Instanton counting via affine Lie algebras I: Equivariant
  J-functions of (affine) flag manifolds and Whittaker vectors}},
  \href{https://arxiv.org/abs/math/0401409}{{\ttfamily math/0401409}}.

\bibitem{Schiffmann:2012aa}
O.~Schiffmann and E.~Vasserot, \emph{{Cherednik algebras, $W$-algebras and the
  equivariant cohomology of the moduli space of instantons on $A^2$}},
  {\emph{Publications math{\'e}matiques de l'IH{\'E}S} {\bfseries 118} (2013)
  213} [\href{https://arxiv.org/abs/1202.2756}{{\ttfamily 1202.2756}}].

\bibitem{Maulik:2012wi}
D.~Maulik and A.~Okounkov, \emph{{Quantum Groups and Quantum Cohomology}},
  {\emph{{Astérisque}} {\bfseries 408} (2019) {ix+209}}
  [\href{https://arxiv.org/abs/1211.1287}{{\ttfamily 1211.1287}}].

\bibitem{Braverman:2014xca}
A.~Braverman, M.~Finkelberg and H.~Nakajima, \emph{{Instanton moduli spaces and
  W-algebras}}, {\emph{{Astérisque}} {\bfseries 385} (2016) vii+128}
  [\href{https://arxiv.org/abs/1406.2381}{{\ttfamily 1406.2381}}].

\bibitem{Atiyah:1978ri}
M.F.~Atiyah, N.J.~Hitchin, V.G.~Drinfeld and Y.I.~Manin, \emph{{Construction of
  Instantons}}, \href{https://doi.org/10.1016/0375-9601(78)90141-X}{\emph{Phys.
  Lett. A} {\bfseries 65} (1978) 185}.

\bibitem{nakajima1994instantons}
H.~Nakajima, \emph{Instantons on {ALE} spaces, quiver varieties, and
  {Kac-Moody} algebras}, {\emph{Duke Mathematical Journal} {\bfseries 76}
  (1994) 365}.

\bibitem{Benvenuti:2010pq}
S.~Benvenuti, A.~Hanany and N.~Mekareeya, \emph{{The Hilbert Series of the One
  Instanton Moduli Space}},
  \href{https://doi.org/10.1007/JHEP06(2010)100}{\emph{JHEP} {\bfseries 1006}
  (2010) 100} [\href{https://arxiv.org/abs/1005.3026}{{\ttfamily 1005.3026}}].

\bibitem{Nekrasov-Shadchin}
N.~Nekrasov and S.~Shadchin, \emph{{ABCD of instantons}},
  \href{https://doi.org/10.1007/s00220-004-1189-1}{\emph{Commun. Math. Phys.}
  {\bfseries 252} (2004) 359}
  [\href{https://arxiv.org/abs/hep-th/0404225}{{\ttfamily hep-th/0404225}}].

\bibitem{Kim:2012gu}
H.-C.~Kim, S.-S.~Kim and K.~Lee, \emph{{5-dim Superconformal Index with
  Enhanced $E_n$ Global Symmetry}},
  \href{https://doi.org/10.1007/JHEP10(2012)142}{\emph{JHEP} {\bfseries 10}
  (2012) 142} [\href{https://arxiv.org/abs/1206.6781}{{\ttfamily 1206.6781}}].

\bibitem{Kim:2018gjo}
H.-C.~Kim, J.~Kim, S.~Kim, K.-H.~Lee and J.~Park, \emph{{6d strings and
  exceptional instantons}},
  \href{https://doi.org/10.1103/PhysRevD.103.025012}{\emph{Phys. Rev. D}
  {\bfseries 103} (2021) 025012}
  [\href{https://arxiv.org/abs/1801.03579}{{\ttfamily 1801.03579}}].

\bibitem{AKMV}
M.~Aganagic, A.~Klemm, M.~Mari\~{n}o and C.~Vafa, \emph{{The Topological
  Vertex}}, \href{https://doi.org/10.1007/s00220-004-1162-z}{\emph{Commun.
  Math. Phys.} {\bfseries 254} (2005) 425}
  [\href{https://arxiv.org/abs/hep-th/0305132}{{\ttfamily hep-th/0305132}}].

\bibitem{Katz:1996fh}
S.H.~Katz, A.~Klemm and C.~Vafa, \emph{{Geometric engineering of quantum field
  theories}}, \href{https://doi.org/10.1016/S0550-3213(97)00282-4}{\emph{Nucl.
  Phys. B} {\bfseries 497} (1997) 173}
  [\href{https://arxiv.org/abs/hep-th/9609239}{{\ttfamily hep-th/9609239}}].

\bibitem{Katz:1997eq}
S.~Katz, P.~Mayr and C.~Vafa, \emph{{Mirror symmetry and exact solution of 4-D
  N=2 gauge theories: 1.}},
  \href{https://doi.org/10.4310/ATMP.1997.v1.n1.a2}{\emph{Adv. Theor. Math.
  Phys.} {\bfseries 1} (1998) 53}
  [\href{https://arxiv.org/abs/hep-th/9706110}{{\ttfamily hep-th/9706110}}].

\bibitem{Leung-Vafa}
N.C.~Leung and C.~Vafa, \emph{{Branes and toric geometry}},
  \href{https://doi.org/10.4310/ATMP.1998.v2.n1.a4}{\emph{Adv. Theor. Math.
  Phys.} {\bfseries 2} (1998) 91}
  [\href{https://arxiv.org/abs/hep-th/9711013}{{\ttfamily hep-th/9711013}}].

\bibitem{Hayashi:2020hhb}
H.~Hayashi and R.-D.~Zhu, \emph{{More on topological vertex formalism for
  5-brane webs with O5-plane}},
  \href{https://doi.org/10.1007/JHEP04(2021)292}{\emph{JHEP} {\bfseries 04}
  (2021) 292} [\href{https://arxiv.org/abs/2012.13303}{{\ttfamily
  2012.13303}}].

\bibitem{G-type}
H.~Hayashi, S.-S.~Kim, K.~Lee and F.~Yagi, \emph{{5-brane webs for 5d $
  \mathcal{N} $ = 1 G$_{2}$ gauge theories}},
  \href{https://doi.org/10.1007/JHEP03(2018)125}{\emph{JHEP} {\bfseries 03}
  (2018) 125} [\href{https://arxiv.org/abs/1801.03916}{{\ttfamily
  1801.03916}}].

\bibitem{Hayashi:2018lyv}
H.~Hayashi, S.-S.~Kim, K.~Lee and F.~Yagi, \emph{{Dualities and 5-brane webs
  for 5d rank 2 SCFTs}},
  \href{https://doi.org/10.1007/JHEP12(2018)016}{\emph{JHEP} {\bfseries 12}
  (2018) 016} [\href{https://arxiv.org/abs/1806.10569}{{\ttfamily
  1806.10569}}].

\bibitem{Shadchin:2005mx}
S.~Shadchin, \emph{{On certain aspects of string theory/gauge theory
  correspondence}},  {Ph.D.} thesis, 2, 2005,
  [\href{https://arxiv.org/abs/hep-th/0502180}{{\ttfamily hep-th/0502180}}].

\bibitem{Bergman:2013ala}
O.~Bergman, D.~Rodr\'\i{}guez-G\'omez and G.~Zafrir, \emph{{Discrete $\theta$
  and the 5d superconformal index}},
  \href{https://doi.org/10.1007/JHEP01(2014)079}{\emph{JHEP} {\bfseries 01}
  (2014) 079} [\href{https://arxiv.org/abs/1310.2150}{{\ttfamily 1310.2150}}].

\bibitem{Kim:2012qf}
H.-C.~Kim, J.~Kim and S.~Kim, \emph{{Instantons on the 5-sphere and
  M5-branes}},  \href{https://arxiv.org/abs/1211.0144}{{\ttfamily 1211.0144}}.

\bibitem{Hwang:2014uwa}
C.~Hwang, J.~Kim, S.~Kim and J.~Park, \emph{{General instanton counting and 5d
  SCFT}}, \href{https://doi.org/10.1007/JHEP07(2015)063}{\emph{JHEP} {\bfseries
  07} (2015) 063} [\href{https://arxiv.org/abs/1406.6793}{{\ttfamily
  1406.6793}}].

\bibitem{Sen:1996vd}
A.~Sen, \emph{{F theory and orientifolds}},
  \href{https://doi.org/10.1016/0550-3213(96)00347-1}{\emph{Nucl. Phys. B}
  {\bfseries 475} (1996) 562}
  [\href{https://arxiv.org/abs/hep-th/9605150}{{\ttfamily hep-th/9605150}}].

\bibitem{Hanany:1996ie}
A.~Hanany and E.~Witten, \emph{{Type IIB superstrings, BPS monopoles, and
  three-dimensional gauge dynamics}},
  \href{https://doi.org/10.1016/S0550-3213(97)00157-0}{\emph{Nucl. Phys. B}
  {\bfseries 492} (1997) 152}
  [\href{https://arxiv.org/abs/hep-th/9611230}{{\ttfamily hep-th/9611230}}].

\bibitem{Hayashi:2016jak}
H.~Hayashi and G.~Zoccarato, \emph{{Partition functions of web diagrams with an
  O7$^{-}$-plane}}, \href{https://doi.org/10.1007/JHEP03(2017)112}{\emph{JHEP}
  {\bfseries 03} (2017) 112}
  [\href{https://arxiv.org/abs/1609.07381}{{\ttfamily 1609.07381}}].

\bibitem{Awata:2005fa}
H.~Awata and H.~Kanno, \emph{{Instanton counting, Macdonald functions and the
  moduli space of D-branes}},
  \href{https://doi.org/10.1088/1126-6708/2005/05/039}{\emph{JHEP} {\bfseries
  05} (2005) 039} [\href{https://arxiv.org/abs/hep-th/0502061}{{\ttfamily
  hep-th/0502061}}].

\bibitem{IKV}
A.~Iqbal, C.~Kozcaz and C.~Vafa, \emph{{The Refined topological vertex}},
  \href{https://doi.org/10.1088/1126-6708/2009/10/069}{\emph{JHEP} {\bfseries
  10} (2009) 069} [\href{https://arxiv.org/abs/hep-th/0701156}{{\ttfamily
  hep-th/0701156}}].

\bibitem{Kim-Yagi}
S.-S.~Kim and F.~Yagi, \emph{{Topological vertex formalism with O5-plane}},
  \href{https://doi.org/10.1103/PhysRevD.97.026011}{\emph{Phys. Rev.}
  {\bfseries D97} (2018) 026011}
  [\href{https://arxiv.org/abs/1709.01928}{{\ttfamily 1709.01928}}].

\bibitem{D-type}
J.-E.~Bourgine, M.~Fukuda, Y.~Matsuo and R.-D.~Zhu, \emph{{Reflection states in
  Ding--Iohara--Miki algebra and brane-web for D-type quiver}},
  \href{https://doi.org/10.1007/JHEP12(2017)015}{\emph{JHEP} {\bfseries 12}
  (2017) 015} [\href{https://arxiv.org/abs/1709.01954}{{\ttfamily
  1709.01954}}].

\bibitem{Ohmori-Hayashi}
H.~Hayashi and K.~Ohmori, \emph{{5d/6d DE instantons from trivalent gluing of
  web diagrams}}, \href{https://doi.org/10.1007/JHEP06(2017)078}{\emph{JHEP}
  {\bfseries 06} (2017) 078}
  [\href{https://arxiv.org/abs/1702.07263}{{\ttfamily 1702.07263}}].

\bibitem{Zafrir:2015ftn}
G.~Zafrir, \emph{{Brane webs and $O5$-planes}},
  \href{https://doi.org/10.1007/JHEP03(2016)109}{\emph{JHEP} {\bfseries 03}
  (2016) 109} [\href{https://arxiv.org/abs/1512.08114}{{\ttfamily
  1512.08114}}].

\bibitem{Bertoldi:2002nn}
G.~Bertoldi, B.~Feng and A.~Hanany, \emph{{The Splitting of branes on
  orientifold planes}},
  \href{https://doi.org/10.1088/1126-6708/2002/04/015}{\emph{JHEP} {\bfseries
  04} (2002) 015} [\href{https://arxiv.org/abs/hep-th/0202090}{{\ttfamily
  hep-th/0202090}}].

\bibitem{Kim:2016qqs}
H.-C.~Kim, \emph{{Line defects and 5d instanton partition functions}},
  \href{https://doi.org/10.1007/JHEP03(2016)199}{\emph{JHEP} {\bfseries 03}
  (2016) 199} [\href{https://arxiv.org/abs/1601.06841}{{\ttfamily
  1601.06841}}].

\bibitem{Haouzi:2020yxy}
N.~Haouzi and J.~Oh, \emph{{On the Quantization of Seiberg-Witten Geometry}},
  \href{https://doi.org/10.1007/JHEP01(2021)184}{\emph{JHEP} {\bfseries 01}
  (2021) 184} [\href{https://arxiv.org/abs/2004.00654}{{\ttfamily
  2004.00654}}].

\bibitem{Nawata-Zhu}
S.~Nawata and R.-D.~Zhu, \emph{{$qq$-characters of BCD types}}, {\emph{to
  appear} }.

\bibitem{Gottsche:2006bm}
L.~Gottsche, H.~Nakajima and K.~Yoshioka, \emph{{K-theoretic Donaldson
  invariants via instanton counting}},
  \href{https://doi.org/10.4310/PAMQ.2009.v5.n3.a5}{\emph{Pure Appl. Math.
  Quart.} {\bfseries 5} (2009) 1029}
  [\href{https://arxiv.org/abs/math/0611945}{{\ttfamily math/0611945}}].

\bibitem{Nakajima:2009qjc}
H.~Nakajima and K.~Yoshioka, \emph{{Perverse coherent sheaves on blowup, III:
  Blow-up formula from wall-crossing}},
  \href{https://doi.org/10.1215/21562261-1214366}{\emph{Kyoto J. Math.}
  {\bfseries 51} (2011) 263} [\href{https://arxiv.org/abs/0911.1773}{{\ttfamily
  0911.1773}}].

\bibitem{Keller:2012da}
C.A.~Keller and J.~Song, \emph{{Counting Exceptional Instantons}},
  \href{https://doi.org/10.1007/JHEP07(2012)085}{\emph{JHEP} {\bfseries 07}
  (2012) 085} [\href{https://arxiv.org/abs/1205.4722}{{\ttfamily 1205.4722}}].

\bibitem{Kim:2019uqw}
J.~Kim, S.-S.~Kim, K.-H.~Lee, K.~Lee and J.~Song, \emph{{Instantons from
  Blow-up}}, \href{https://doi.org/10.1007/JHEP11(2019)092}{\emph{JHEP}
  {\bfseries 11} (2019) 092}
  [\href{https://arxiv.org/abs/1908.11276}{{\ttfamily 1908.11276}}].

\bibitem{Huang:2017mis}
M.-x.~Huang, K.~Sun and X.~Wang, \emph{{Blowup Equations for Refined
  Topological Strings}},
  \href{https://doi.org/10.1007/JHEP10(2018)196}{\emph{JHEP} {\bfseries 10}
  (2018) 196} [\href{https://arxiv.org/abs/1711.09884}{{\ttfamily
  1711.09884}}].

\bibitem{Hollands:2010xa}
L.~Hollands, C.A.~Keller and J.~Song, \emph{{From SO/Sp instantons to W-algebra
  blocks}}, \href{https://doi.org/10.1007/JHEP03(2011)053}{\emph{JHEP}
  {\bfseries 03} (2011) 053} [\href{https://arxiv.org/abs/1012.4468}{{\ttfamily
  1012.4468}}].

\bibitem{Hollands:2011zc}
L.~Hollands, C.A.~Keller and J.~Song, \emph{{Towards a 4d/2d correspondence for
  Sicilian quivers}},
  \href{https://doi.org/10.1007/JHEP10(2011)100}{\emph{JHEP} {\bfseries 1110}
  (2011) 100} [\href{https://arxiv.org/abs/1107.0973}{{\ttfamily 1107.0973}}].

\bibitem{Keller:2011ek}
C.A.~Keller, N.~Mekareeya, J.~Song and Y.~Tachikawa, \emph{{The ABCDEFG of
  Instantons and W-algebras}},
  \href{https://doi.org/10.1007/JHEP03(2012)045}{\emph{JHEP} {\bfseries 03}
  (2012) 045} [\href{https://arxiv.org/abs/1111.5624}{{\ttfamily 1111.5624}}].

\bibitem{Gamayun:2012ma}
O.~Gamayun, N.~Iorgov and O.~Lisovyy, \emph{{Conformal field theory of
  Painlev\'e VI}}, \href{https://doi.org/10.1007/JHEP10(2012)038}{\emph{JHEP}
  {\bfseries 10} (2012) 038} [\href{https://arxiv.org/abs/1207.0787}{{\ttfamily
  1207.0787}}].

\bibitem{Nekrasov:2020qcq}
N.~Nekrasov, \emph{{Blowups in BPS/CFT correspondence, and Painlev\'e VI}},
  \href{https://arxiv.org/abs/2007.03646}{{\ttfamily 2007.03646}}.

\bibitem{Jeong:2020uxz}
S.~Jeong and N.~Nekrasov, \emph{{Riemann-Hilbert correspondence and blown up
  surface defects}}, \href{https://doi.org/10.1007/JHEP12(2020)006}{\emph{JHEP}
  {\bfseries 12} (2020) 006}
  [\href{https://arxiv.org/abs/2007.03660}{{\ttfamily 2007.03660}}].

\bibitem{Bonelli:2021rrg}
G.~Bonelli, F.~Globlek and A.~Tanzini, \emph{{Counting Yang-Mills Instantons by
  Surface Operator Renormalization Group Flow}},
  \href{https://doi.org/10.1103/PhysRevLett.126.231602}{\emph{Phys. Rev. Lett.}
  {\bfseries 126} (2021) 231602}
  [\href{https://arxiv.org/abs/2102.01627}{{\ttfamily 2102.01627}}].

\bibitem{Gopakumar:1998ki}
R.~Gopakumar and C.~Vafa, \emph{{On the gauge theory / geometry
  correspondence}},
  \href{https://doi.org/10.4310/ATMP.1999.v3.n5.a5}{\emph{AMS/IP Stud. Adv.
  Math.} {\bfseries 23} (2001) 45}
  [\href{https://arxiv.org/abs/hep-th/9811131}{{\ttfamily hep-th/9811131}}].

\bibitem{Marino:2009mw}
M.~Marino, \emph{{String theory and the Kauffman polynomial}},
  \href{https://doi.org/10.1007/s00220-010-1088-6}{\emph{Commun. Math. Phys.}
  {\bfseries 298} (2010) 613}
  [\href{https://arxiv.org/abs/0904.1088}{{\ttfamily 0904.1088}}].

\bibitem{Okounkov:2003sp}
A.~Okounkov, N.~Reshetikhin and C.~Vafa, \emph{{Quantum Calabi-Yau and
  classical crystals}},
  \href{https://doi.org/10.1007/0-8176-4467-9_16}{\emph{Prog. Math.} {\bfseries
  244} (2006) 597} [\href{https://arxiv.org/abs/hep-th/0309208}{{\ttfamily
  hep-th/0309208}}].

\bibitem{Nekrasov:2003rj}
N.~Nekrasov and A.~Okounkov, \emph{{Seiberg-Witten theory and random
  partitions}}, \href{https://doi.org/10.1007/0-8176-4467-9_15}{\emph{Prog.
  Math.} {\bfseries 244} (2006) 525}
  [\href{https://arxiv.org/abs/hep-th/0306238}{{\ttfamily hep-th/0306238}}].

\bibitem{Macdonald-book}
I.G.~Macdonald, \emph{{Symmetric Functions and Hall Polynomials}}, OXFORD
  SCIENCE PUBLICATIONS (1979).

\bibitem{jeffrey1995localization}
L.C.~Jeffrey and F.C.~Kirwan, \emph{Localization for nonabelian group actions},
  {\emph{Topology} {\bfseries 34} (1995) 291}
  [\href{https://arxiv.org/abs/alg-geom/9307001}{{\ttfamily
  alg-geom/9307001}}].

\bibitem{Benini:2013xpa}
F.~Benini, R.~Eager, K.~Hori and Y.~Tachikawa, \emph{{Elliptic Genera of 2d
  ${\mathcal{N}}$ = 2 Gauge Theories}},
  \href{https://doi.org/10.1007/s00220-014-2210-y}{\emph{Commun. Math. Phys.}
  {\bfseries 333} (2015) 1241}
  [\href{https://arxiv.org/abs/1308.4896}{{\ttfamily 1308.4896}}].

\bibitem{Marino:2004cn}
M.~Mari\~{n}o and N.~Wyllard, \emph{{A Note on instanton counting for
  $\mathcal{N}=2$ gauge theories with classical gauge groups}},
  \href{https://doi.org/10.1088/1126-6708/2004/05/021}{\emph{JHEP} {\bfseries
  05} (2004) 021} [\href{https://arxiv.org/abs/hep-th/0404125}{{\ttfamily
  hep-th/0404125}}].

\bibitem{Fucito:2004gi}
F.~Fucito, J.F.~Morales and R.~Poghossian, \emph{{Instantons on quivers and
  orientifolds}},
  \href{https://doi.org/10.1088/1126-6708/2004/10/037}{\emph{JHEP} {\bfseries
  10} (2004) 037} [\href{https://arxiv.org/abs/hep-th/0408090}{{\ttfamily
  hep-th/0408090}}].

\bibitem{Nakamura:2014nha}
S.~Nakamura, F.~Okazawa and Y.~Matsuo, \emph{{Recursive method for the Nekrasov
  partition function for classical Lie groups}},
  \href{https://doi.org/10.1093/ptep/ptv014}{\emph{PTEP} {\bfseries 2015}
  (2015) 033B01} [\href{https://arxiv.org/abs/1411.4222}{{\ttfamily
  1411.4222}}].

\bibitem{Nakamura:2015zsa}
S.~Nakamura, \emph{{On the Jeffrey\textendash{}Kirwan residue of
  BCD-instantons}}, \href{https://doi.org/10.1093/ptep/ptv085}{\emph{PTEP}
  {\bfseries 2015} (2015) 073B02}
  [\href{https://arxiv.org/abs/1502.04188}{{\ttfamily 1502.04188}}].

\bibitem{Hwang:2017nop}
C.~Hwang and P.~Yi, \emph{{Twisted Partition Functions and $H$-Saddles}},
  \href{https://doi.org/10.1007/JHEP06(2017)045}{\emph{JHEP} {\bfseries 06}
  (2017) 045} [\href{https://arxiv.org/abs/1704.08285}{{\ttfamily
  1704.08285}}].

\bibitem{Hwang:2018riu}
C.~Hwang, S.~Lee and P.~Yi, \emph{{Holonomy Saddles and Supersymmetry}},
  \href{https://doi.org/10.1103/PhysRevD.97.125013}{\emph{Phys. Rev. D}
  {\bfseries 97} (2018) 125013}
  [\href{https://arxiv.org/abs/1801.05460}{{\ttfamily 1801.05460}}].

\end{thebibliography}\endgroup
\end{document}